\documentclass{aa}  
\usepackage{natbib}
\bibpunct{(}{)}{;}{a}{}{,} % follow the A&A style
\usepackage{graphicx}
\usepackage{txfonts}
\usepackage{hyperref}
\graphicspath{{./}{plot_paper/}}
\begin{document} 

\title{Ongoing hierarchical massive cluster assembly: the LISCA~II structure in the Perseus complex}

\subtitle{}
\author{A. Della Croce\inst{1,2}\thanks{\email{alessandro.dellacroce@inaf.it}}
      \and
      E. Dalessandro\inst{2}
      \and 
      A. Livernois\inst{3}
      \and 
      E. Vesperini\inst{3}
      \and
      C. Fanelli\inst{2}
      \and
      L. Origlia\inst{2}
      \and 
      M. Bellazzini\inst{2}
      \and
      E. Oliva\inst{4}
      \and
      N. Sanna\inst{4}
      \and
      A. L. Varri\inst{5}
}

\institute{
    Department of Physics and Astronomy ‘Augusto Righi’, University of Bologna, via Gobetti 93/2, I-40129 Bologna, Italy
    \and
    INAF – Astrophysics and Space Science Observatory of Bologna, via Gobetti 93/3, I-40129 Bologna, Italy
    \and 
    Department of Astronomy, Indiana University, Swain West, 727 E. 3rd Street, IN 47405 Bloomington, USA
    \and 
    INAF - Osservatorio Astrofisico di Arcetri, Largo Enrico Fermi 5, I-50125 Florence, Italy
    \and 
    Institute for Astronomy, University of Edinburgh, Royal Observatory, Blackford Hill, Edinburgh EH9 3HJ, UK
}

   \date{\today}

\abstract 
{
We report on the identification of a massive  ($\sim10^5$ M$_\odot$) sub-structured stellar system in the Galactic Perseus complex likely undergoing hierarchical cluster assembly. Such a system comprises nine star clusters (including the well-known clusters NGC 654 and NGC 663) and an extended and low-density stellar halo. Gaia-DR3 and available spectroscopic data show that all its components are physically consistent in the 6D phase-space (position, parallax, and 3D motion), homogeneous in age (14 $-$ 44 Myr), and chemical content (half-solar metallicity). 
In addition, the system's global stellar density distribution is that of typical star clusters and shows clear evidence
of mass segregation. We find that the hierarchical structure is mostly contracting towards the center with a speed of up to  $\simeq4-5$ km s$^{-1}$, while the innermost regions expand at a lower rate (about $\simeq1$ km s$^{-1}$) and are dominated by random motions. 
Interestingly, this pattern is dominated by the kinematics of massive stars, while low-mass stars ($M<2$ M$_\odot$) are characterized by contraction across the entire cluster. 
Finally, the nine star clusters in the system are all characterized by a relatively flat velocity dispersion profile possibly resulting from ongoing interactions and tidal heating.
We show that the observational results are generally consistent with those found in $N$-body simulations following the cluster violent relaxation phase strongly suggesting that the system is a massive cluster in the early assembly stages.
This is the second structure with these properties identified in our Galaxy and, following the nomenclature of our previous work, we named it LISCA II.
}


\maketitle

\section{Introduction} \label{sec:intro}
Cluster formation is important for the study of many key questions in modern astrophysics.
Firstly, it is central to the star-formation process itself \citep[e.g.][]{mckee_ostriker_2007}. In fact, it is commonly accepted that most ($70\%-90\%$) stars in galaxies form in groups, 
clusters, or hierarchical systems, and spend some time gravitationally bound with their siblings 
when still embedded in their progenitor molecular cloud  \citep{lada&lada2003}. 
In the Milky Way (MW), this evidence comes from the global clustered structure of the disc and spiral arms \citep[e.g.][]{kounkel&covey2019} and from the similar star formation rates observed in embedded clusters 
\citep{lada&lada2003} and in the field \citep{miller&scalo1979}. 
On a larger scale, this indication is supported by the good agreement between the mass 
density in stellar clusters and the average comoving stellar density at the peak of the 
universal star formation density at redshift $\sim2$ \citep{madau&dickinson2014}.

Secondly, cluster formation has many implications for the early interplay between stellar and gas dynamics, 
the possible formation of gravitational wave sources 
\citep{dicarlo+19,Banerjee2021} and the dynamical properties of young star clusters \citep[e.g.][]{mcmillan+2007,ballone+20,ballone+21,livernois+2021,tiongco+22}.
Finally, cluster formation is fundamental for our understanding of the assembly process of galaxies in a cosmological context, as major star-forming episodes in galaxies are typically accompanied 
by significant star cluster production \citep{forbes+18} and the main properties of these systems are thus strictly 
connected with those of their hosts \citep{brodie_strader_2006, dalessandro+12}. 

However, the possible presence of unifying principles governing the formation of stellar clusters and whether they form
through a monolithic event or as a result of a hierarchical process of star formation in which stars 
are formed across a continuous distribution of gas densities, is still a matter of debate \citep{lada+1984,moeckel&bonnell2009, kruijssen2012,banerjee&kroupa2014,banerjee&kroupa2015,trevino-morales+19,kuhn+2019,pang+2022}.
In addition, the observed presence in the most massive clusters of the so-called multiple stellar populations characterized 
by specific abundance patterns in a number of light elements (see for example \citealt{bsatian_lardo_2018,Gratton+19} for recent reviews) 
has raised new questions on the physical mechanisms at the basis of clusters' formation, 
and their dependence on the environment and the formation epoch \citep{Krumholz+19}.
As a matter of fact, despite the tremendous observational and theoretical efforts in the recent years 
\citep{allison_etal2010,parker2014,adamo_etal2015},
our understanding of how and where star clusters form is still in its infancy.

Numerous recent observational studies of stellar clusters have significantly enriched our knowledge about the properties of 
these systems. Indeed, the unprecedented kinematical mapping of the Galaxy and its stellar components,
as secured by the Gaia mission 
\citep{gaiaEDR3_2020}, 
is revolutionizing the field \citep{cantat-gaudin&anders+2020,cantat-gaudin+2020,castro-ginard+2022,cantat-gaudin_reviewGaia} enabling detailed studies 
of nearby star-forming regions and their use as ideal laboratories 
to shed new light on our understanding of cluster formation and early evolution
\citep[e.g.][]{beccari_etal2018,meigast_etal2019,lim_etal2020}.
Indeed, observations have revealed significant structural and kinematical complexity such as, for instance, significant deviations from spherical symmetry, the presence of extended tails in both young
\citep[e.g.][]{meingast+21,jerabkova+21,pang+21} and old star clusters
\citep[e.g.][]{grillmair+2019,bonaca+2020},
as well as evidence of internal rotation and/or radially anisotropic velocity distributions \citep[e.g.][]{henault-brunet+2012,ferraro+18,kamann+18,vasiliev_baumgardt_21,dalessandr+21-GCs}.   

It has also been shown that young stellar systems have often a complex clumpy structure characterized by the presence of several stellar subsystems \citep[e.g.][]{kuhn+2019,getman+19,kuhn+20,lim_etal2020,dalessandro+2021,zeidler+21}.
While some of these systems are likely to dissolve, some may evolve into massive and long-lived clusters.
Interestingly in this context, \citet{dalessandro+2021} have found that the well-known clusters $h$ and $\chi$ Persei are just components of an association of clusters embedded in a wide stellar halo of similar age. This structure, named LISCA~I,
has provided the first detailed observational picture of an ongoing massive cluster hierarchical assembly.  
This is the first time that such a formation mechanism has been identified in the MW
and it has important implications 
on our understanding of the environmental conditions 
(both locally and in the distant Universe) necessary to form massive stellar clusters. 
For many years, hierarchical cluster formation has been invoked as the preferred dynamical route 
to form rotating and high ellipticity star clusters 
\citep[e.g.][]{deOliveira+1998}. 
More recently, it has been used to interpret the properties of particularly massive and 
dynamically complex clusters 
\citep[e.g.][]{lee+1999,bruns_kroupa_2011}, and as an avenue 
to form clusters with multiple populations with different light elements abundances
\citep[e.g.][]{gavagnin+16,hong+17}.

However, while hierarchical cluster formation is  
believed to work efficiently in high-density starburst 
galaxies 
\citep[e.g.][]{bastian+11,chandar+11}, 
we are still missing an adequate understanding of its effectiveness in lower density 
environments, like the MW and the Magellanic Clouds.

Dynamical simulations \citep{bonnell+2003,ballone+20,livernois+2021} show that within a hierarchical assembly framework, 
the fragmentation of a molecular cloud 
may lead firstly to the formation of tens
of small clumps ($\sim100 M_{\odot}$), then the surviving clumps merge to form a 
few more massive ($10^3 - 10^4 M_{\odot}$) and larger clusters. Finally, 
one or two clusters survive this hierarchical merger 
process and will eventually evolve into a single massive cluster
\citep[e.g.][]{fujii_portegeisZwart_2016,livernois+2021}.

As a part of a larger project aimed at constraining the occurrence of the hierarchical assembly process 
within local disk-like galaxies, and 
test whether it is able to form long-lived systems surviving the initial and turbulent 
few tens of million years of existence 
\citep[e.g.][]{moeckel_bate_2010,gieles+06}, we present a detailed photometric and kinematic study, mainly based on Gaia DR3 data \citep{gaiaDR3_2022}, of a region in the Galactic Perseus Arm including the clusters NGC~663 and NGC~654, that appears to be analogous to LISCA~I.
We also present a comparison of the observational results with the dynamical properties emerging in $N$-body models following the violent relaxation phase of a stellar system and its subsequent evolution.

The paper is structured as follows.
The adopted data set is presented in Section \ref{sec:catalog_and targetClusters}; Sections \ref{sec:coherence_physical_prop} and \ref{sec:kin_struct} describe the physical properties of the area under study, its structure and kinematic respectively. 
Section \ref{sec:star_clusters} presents the physical properties of star clusters belonging to the system, while in Section \ref{sec:total_system_mass} we discuss the total system's mass.
A comparison with a set of {\it N}-body simulations 
is described in Section \ref{sec:simulation}. Finally, the main conclusions are drawn in Section \ref{sec:conclusion}.

\section{Catalogs and preliminary analyses} \label{sec:catalog_and targetClusters}

\subsection{The catalogs}\label{sec:catalog}
From the Gaia Archive\footnote{\url{https://gea.esac.esa.int/archive/}.} we retrieved DR3 data for sources distributed within a large area on the sky (5$^\circ$ in radius) arbitrarily centered on the position of NGC~654 and having \emph{five-parameters} astrometric solution (i.e. sources with sky position, proper motion, and parallax measurements) and $G<19.5$ mag. Such a catalog comprised 4.5 million sources.

We supplemented this data set with high-resolution optical and 
near-infrared spectra obtained with the HARPS-N \citep{cosentino+14} and GIANO-B \citep{oliva+12,tozzi+16}
spectrographs at the TNG as part of the 
 \emph{SPA - Stellar Population Astrophysics: the detailed, age-resolved chemistry of the Milky Way disk} Large Program
(Program ID A37TAC13, PI: L. Origlia).
Line-of-sight (LOS) velocities have been obtained for all the observed stars, while detailed chemical abundances for the sub-sample of red supergiants have been computed by \citet{fanelli_etal2022}.

\subsection{Clustering Analysis}\label{sec:clustering_analysis}
As any coherent stellar structure in the considered area should appear as an overdensity in the multi-dimensional phase-space of positions and velocities, 
we performed a clustering analysis on the whole catalog by means of the Hierarchical Density-Based Spatial Clustering of Application with Noise 
(\texttt{HDBSCAN}) algorithm \citep{hdbscan}.
For each star, we used as inputs the galactic coordinates, 
parallax, and proper motion components $(\ell, b, \varpi, \mu_{\alpha *}, \mu_\delta)$, and we set 
the \texttt{HDBSCAN} parameters as \texttt{min\_cluster\_size} = 40 and \texttt{min\_samples} = 30. 
\texttt{min\_cluster\_size} sets a lower limit to the number of objects an overdensity should have to be identified as a cluster (hence we could not identify clusters with less than 40 members), while 
\texttt{min\_samples} represents the number of sources used in determining the nearest neighbor distance for each source. Hence, increasing \texttt{min\_samples} will increase the \emph{mutual reachability distance} among sources and only the densest areas survive as clusters\footnote{we refer to the online documentation (\url{https://hdbscan.readthedocs.io/en/latest/index.html}) for further details.}.
Furthermore, \texttt{HDBSCAN} assigns a cluster membership probability to each star based on its distance from the neighboring stars. The closer the star is to the other cluster's members, the higher the membership probability and vice-versa.

We identified 131 clustered systems within the full 5$^\circ$-wide field of view.
To exclude spurious detections and select only systems that can be classified as clusters with high significance level, 
we followed the post-processing approach described by \citet{hunt&reffert2021}, which uses the nearest-neighbors distance
as a proxy for the local density.
Only structures with a median value of the nearest-neighbors distance smaller than the one of field stars 
at a $3\sigma$ level according to a Mann-Whitney statistics \citep{mann_withney1947} were flagged as true stellar clusters.
Out of 131 putative clusters, 54 systems fulfilled these criteria and were retained for the subsequent analysis. 
Recent open clusters catalogs \citep[e.g.][]{cantat-gaudin&anders+2020,cantat-gaudin+2020,castro-ginard+2022} list 45 clusters in the region with more than 40 members (that is our threshold for identification). Interestingly, we recovered all the known clusters but two (hence 11 unknown structures have been identified by this study), namely UBC~186 and UPK~265.
We verified that UPK~265 could have been recovered by slightly changing the input parameters we set for the clustering analysis, however, it would have been excluded by the preliminary parallax selection \citep[according to its value reported by][]{cantat-gaudin+2020} described below.
The case of UBC~186 is a more interesting one.  A careful investigation of its members reveals significant overlap with 
NGC~581 \citep[128 out of 131 of NGC~581 members are in fact in common with UBC~186,][]{cantat-gaudin+2020}. 
Indeed, our analysis was able to properly identify both NGC~581 and another nearby structure that was labeled as UBC~186 by \citet{cantat-gaudin+2020}. However, the latter was flagged as a false detection by the adopted post-processing routine.
It is important to stress here, however, that the following analysis and the results of this paper do not depend on the inclusion/exclusion of any specific sub-structure or cluster.

Starting from the sample of 54 structures, we performed a preliminary selection to identify clusters sharing 3D position and 2D velocity with NGC 654 \citep[$\varpi = 0.31\pm0.05$ mas, $\mu_{\alpha *} = -1.1\pm0.1$ mas yr$^{-1}$ and $\mu_{\delta} = -0.3\pm0.1$ mas yr$^{-1}$, obtained by][using Gaia DR2 data]{cantat-gaudin+2020}, retaining only those with distance $D = 2.8-3.2$ kpc and co-moving within about 5.5 km s$^{-1}$ (corresponding to $0.38$ mas yr$^{-1}$ at 3 kpc),
according to their median parallax and proper motion estimated from Gaia DR3 data. 
Nine clusters (comprising NGC 654 itself) were selected that way.
We note in passing that none of the 11 previously unknown structures fulfilled these criteria.

\begin{figure}[h!]
\centering
\includegraphics[width=0.5\textwidth]{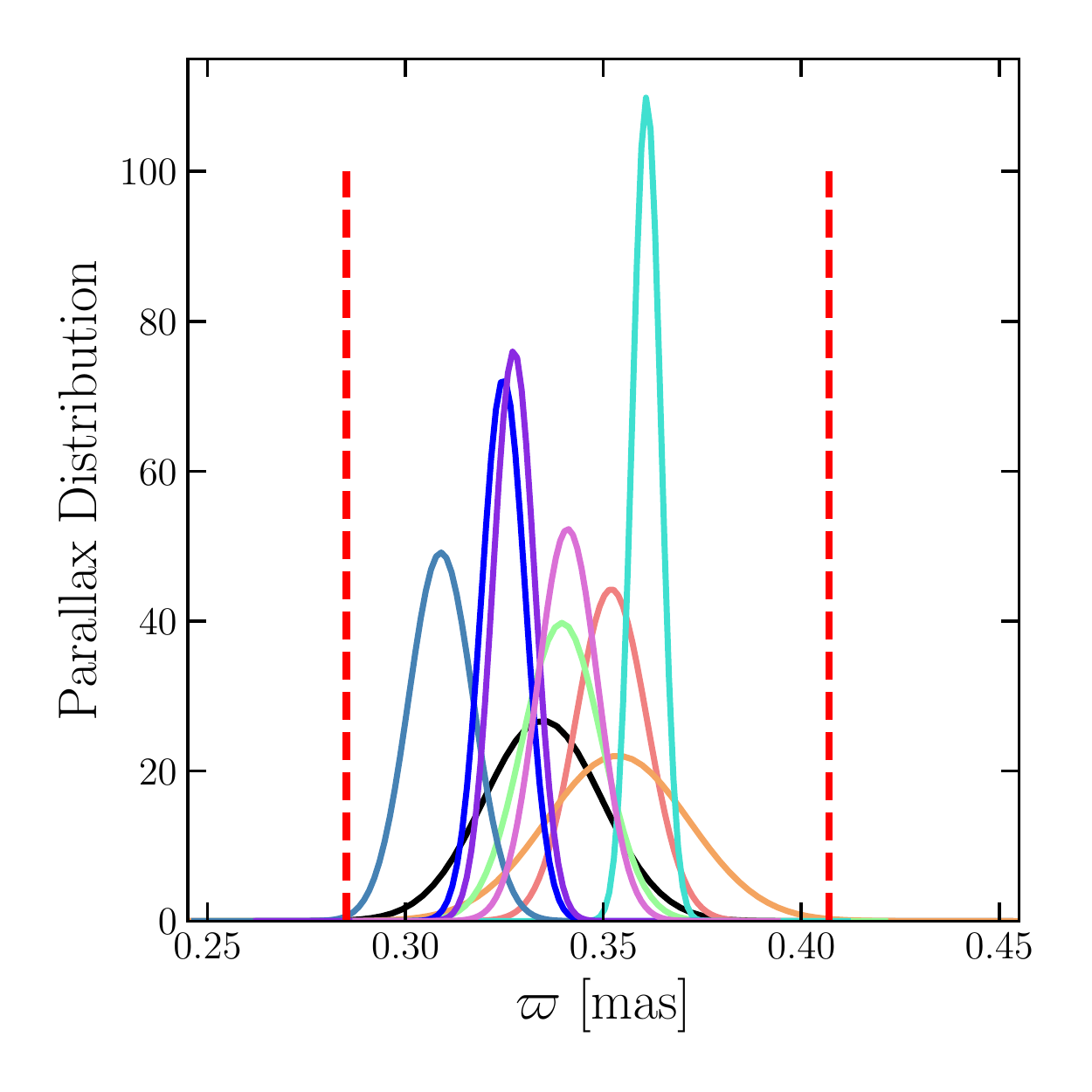}
\includegraphics[width=0.5\textwidth]{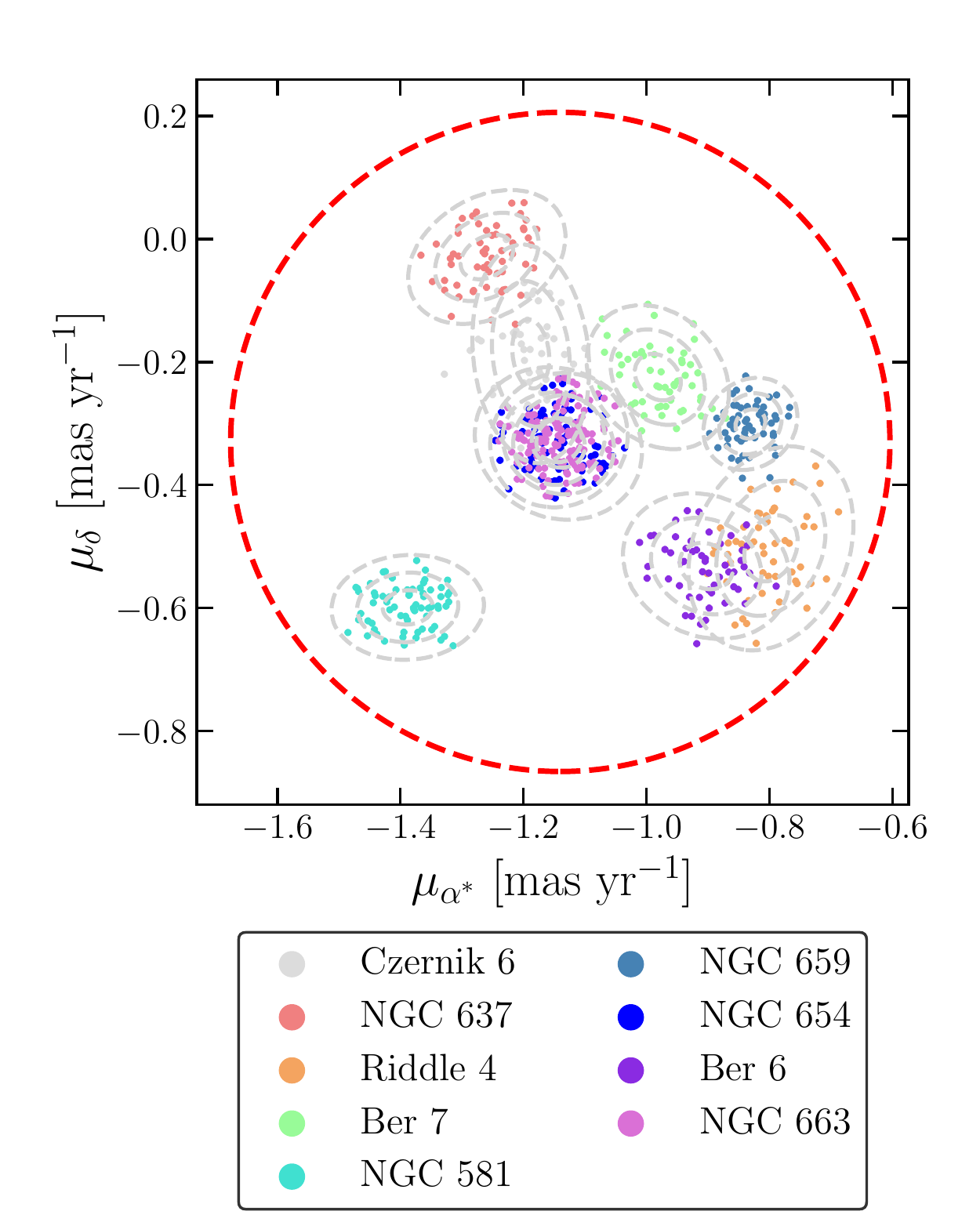}
\caption{Inferred distributions in parallax (top panel) and proper motion (bottom panel) from likely ($>90\%$) cluster members. Different stellar clusters are in different colors. Dashed gray lines in the bottom panel are iso-probability contours at the $1$, $2$, and $3\sigma$ levels respectively. The red dashed circle and vertical lines represent the range in proper motion and parallax inside which stars have been selected.  \label{fig:region_selection}}
\end{figure}

Finally, we determined physically-motivated selections in parallax and proper motion with the aim of selecting all the sources in the field of view sharing 3D position and 2D velocity with the nine clusters. 
Specifically, we inferred the intrinsic clusters' distributions in parallax and proper motion (by using only stars with membership probability larger than 90\%) by means of a Gaussian mixture modeling technique \citep[we used the Extreme Deconvolution\footnote{\url{https://github.com/jobovy/extreme-deconvolution}} package developed by][]{JoBovy_XDGMM}, thereby properly accounting for errors and correlation between measurements.
In Figure \ref{fig:region_selection} we show the distributions inferred that way in both parallax (top panel) 
and proper motion components (bottom panel).

\begin{figure}[h!]
    \centering
    \includegraphics[width=0.5\textwidth]{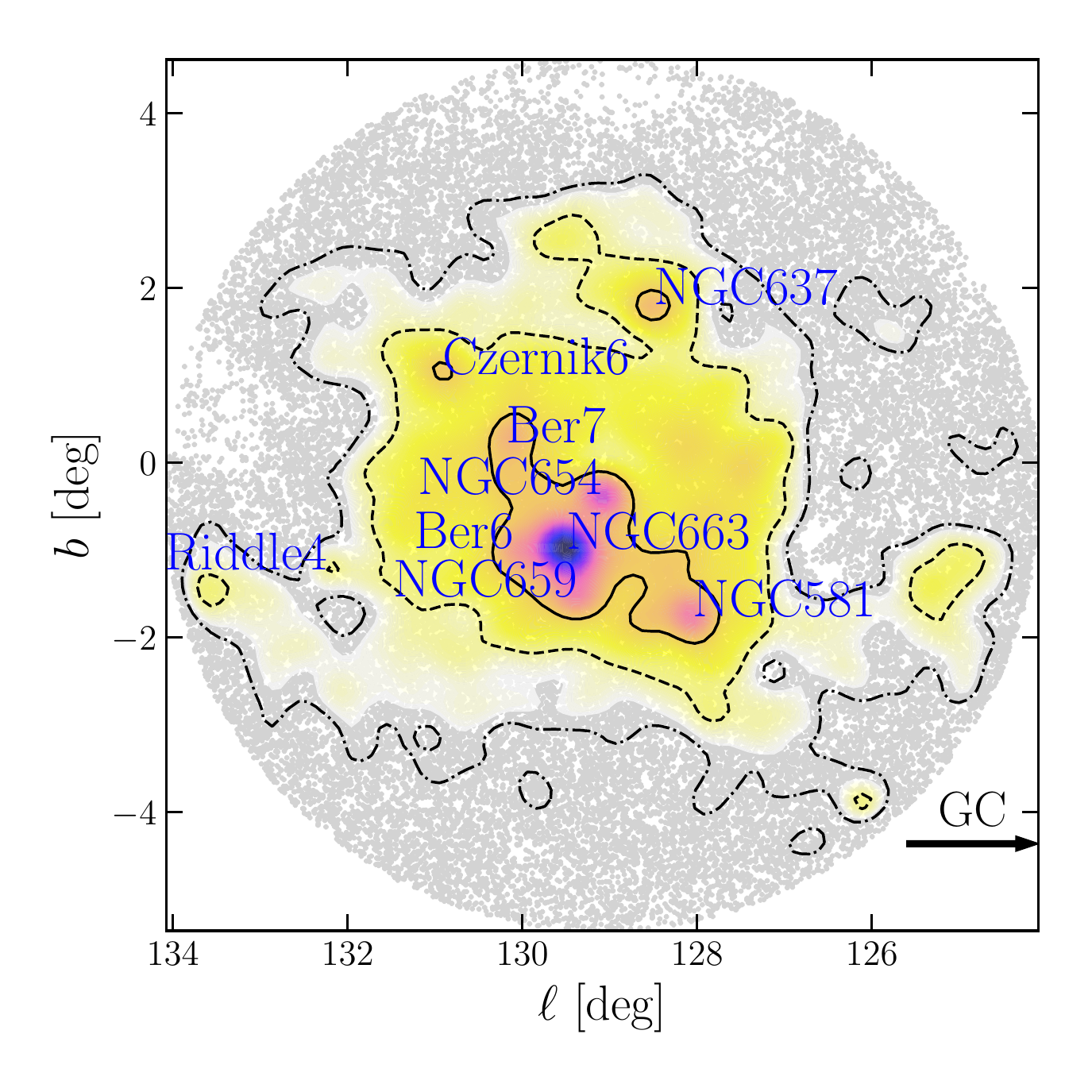}
    \caption{Spatial distribution in Galactic coordinates of stars selected in proper motion and parallax. Star clusters' names are shown in blue, while black lines are iso-density contours enclosing the 11.8\% (solid), 39.3\% (dashed), and 67.5\% (dash-dotted) of the normalized star density distribution. The black arrow shows the direction of the galactic center (GC). \label{fig:fov}}
\end{figure}

We thus retained all the sources in the \emph{Gaia} catalog with proper motions and parallaxes compatible, within $3\sigma$, to 
the clusters' distribution.
Selected stars share similar distances $\varpi\,\in\,[0.285;\,0.407]$ mas, which corresponds to $D\,\in\,[2.46;\,3.51]$ kpc, and
co-move within $0.536$ mas yr$^{-1}$ (about $7.5$ km s$^{-1}$).
In Figure \ref{fig:fov} we show the 2D density map of the region along with iso-density contours.
The iso-density curves highlight the presence of small-scale, clumpy structures corresponding to the identified stellar clusters (labeled in blue) as well as a lower-density diffuse halo extending for at least $3^\circ$ from NGC~663 and NGC~654 and comoving with the clusters.

\subsection{Completeness of the Gaia catalog} \label{sec:completeness}
The estimate of Gaia's catalog completeness is certainly a challenge due to, for instance, a composite data reduction pipeline and a complex satellite scanning law. Moreover, it does not depend only on the telescope properties themselves but also on the physical properties of the observed regions such as crowding and extinction.
This issue has been first tackled by \citet[][for the DR2]{everall&boubert2021-DR2} and \citet[][for the EDR3]{everall&boubert2022-EDR3} who directly modeled Gaia's reduction pipeline and scanning law. More recently, \citet{catant-gaudin22_gaiaUnlimited} adopted an empirical approach to estimate the photometric completeness in the $G-$band by comparing the Gaia catalog with the Dark Energy Camera Plane Survey \citep{schlafly_etal2018,saydjari_etal2022}.

\begin{figure}[h!]
    \centering
    \includegraphics[width=0.5\textwidth]{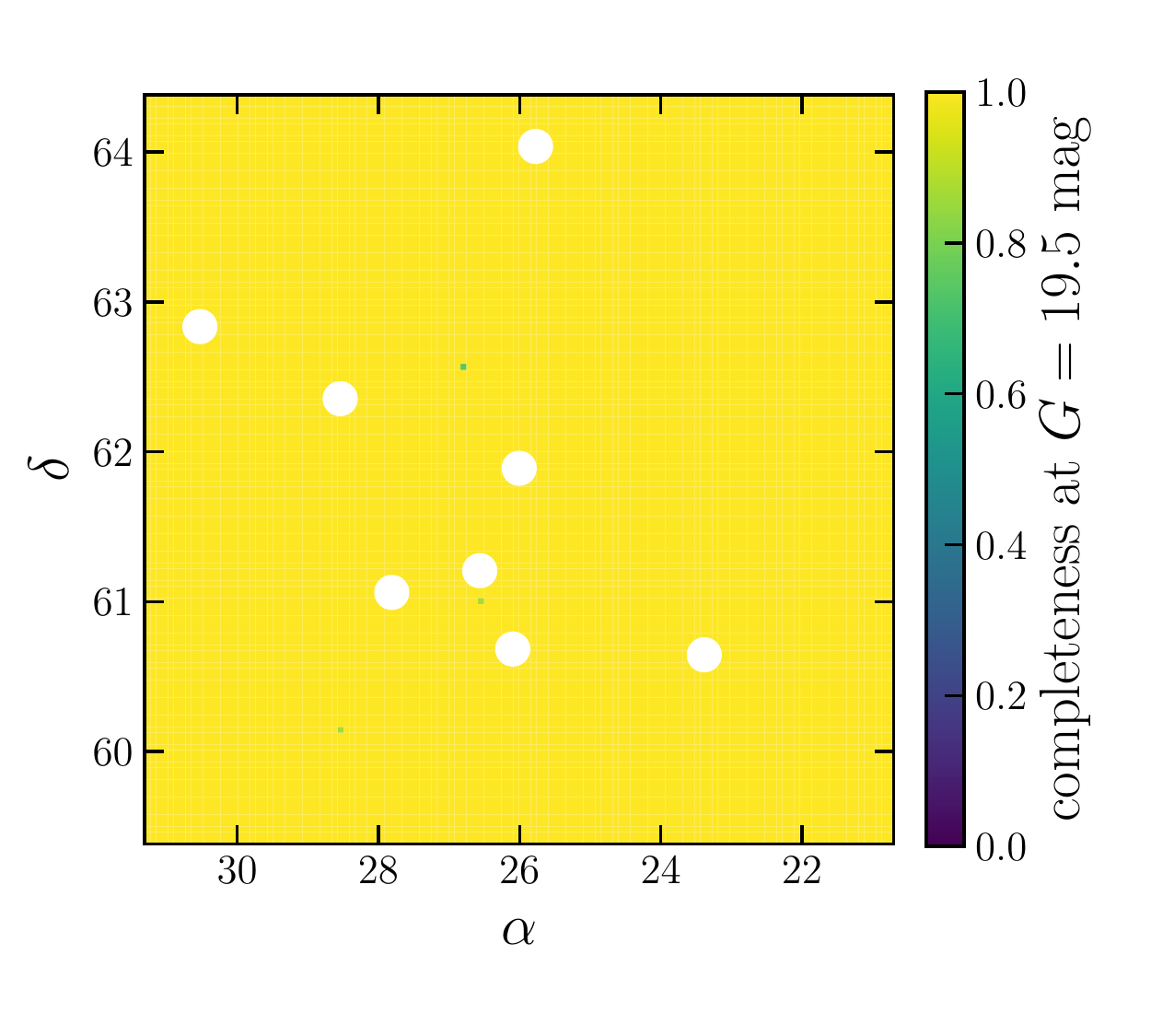}
    \caption{Two-dimensional map of the $p(G)$ term in Equation \ref{eq:completeness} for $G=19.5$ mag, computed with the GaiaUnlimited package \citep{catant-gaudin22_gaiaUnlimited}. White points show the location of star clusters. \label{fig:completeness-Gband}}
\end{figure}

For the purpose of this study, we need to assess the probability that a source is included in the catalog with magnitude $G$ measure and 5 parameters solution.
The resulting joint probability is
\begin{equation}
    p({\rm 5\,params}\,\cdot\,G) = p(G)\times p({\rm 5\,params}\,|\,G)\,.
    \label{eq:completeness}
\end{equation}
We retrieved the first term on the right-hand side from the completeness maps of \citet{catant-gaudin22_gaiaUnlimited}, see Figure \ref{fig:completeness-Gband} for the two-dimensional map of photometric completeness computed for $G=19.5$ mag, whereas the latter was estimated from the number count ratios between sources with 5 parameters solution ($k$) compared to the total number of sources ($n$) for a given sky patch and magnitude bin
\begin{equation}
    p({\rm 5\,params}\,|\,G) = \frac{k+1}{n+2}\,,
    \label{eq:completeness-5params}
\end{equation}
following the documentation of the GaiaUnlimited project\footnote{see for instance \url{https://gaiaunlimited.readthedocs.io/en/latest/dr3-rvs.html}}.
We computed Equation \ref{eq:completeness-5params} for a regular spatial grid in a region 10$^\circ$ wide around NGC 654 assuming a spatial bin size of $\Delta\delta = 0.2^\circ$ (thus it follows $\Delta\alpha = \Delta\delta\,\cos\delta_{\rm NGC654}\simeq0.42^\circ$ in order to obtain a square grid) and for magnitude bins $\Delta G = 0.2$ mag wide down to $G \leq 19.5$ mag.
\begin{figure*}[h!]
    \centering
    \begin{minipage}{0.475\textwidth}
        \includegraphics[width=\textwidth]{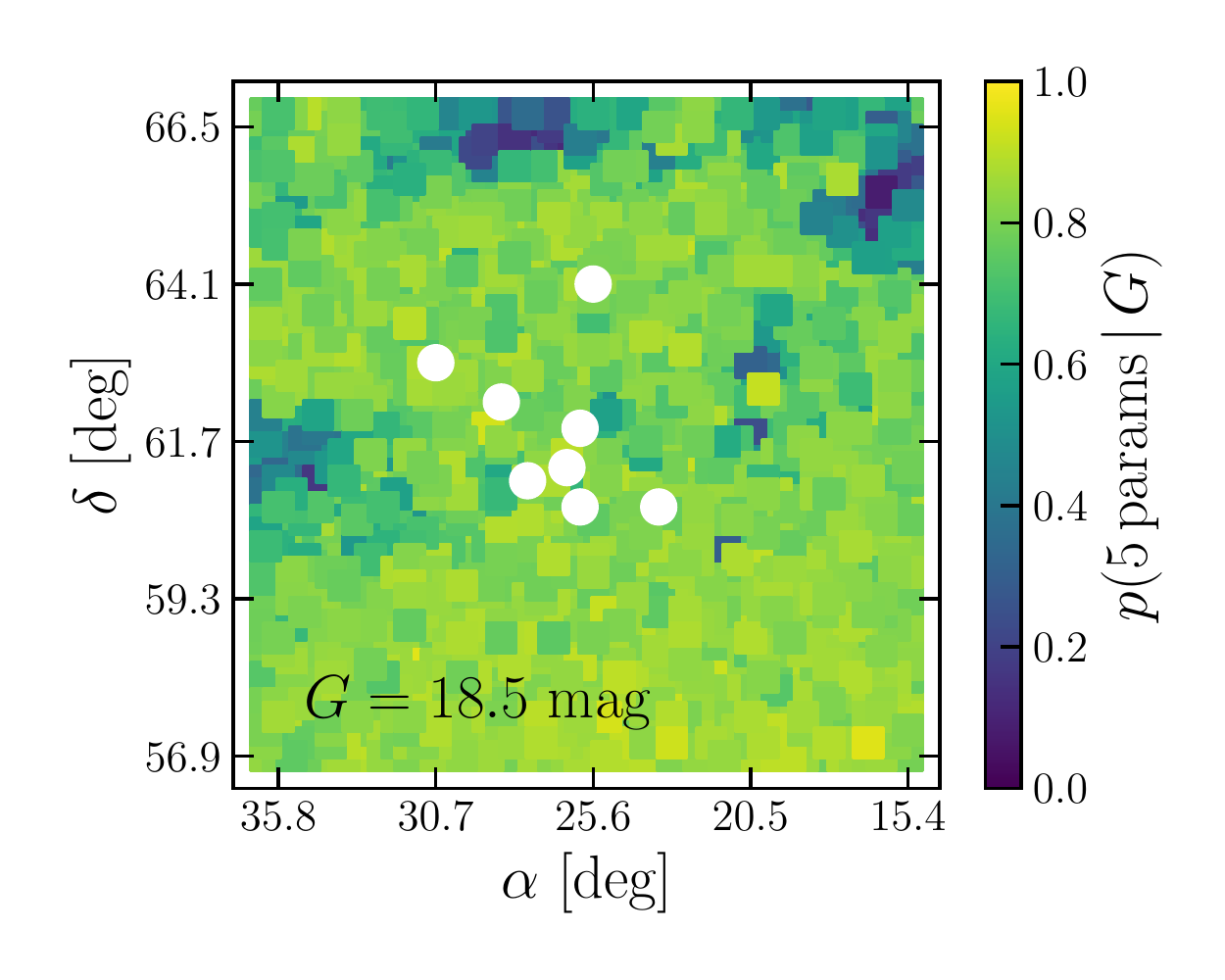}
    \end{minipage}
    \begin{minipage}{0.475\textwidth}
        \includegraphics[width=\textwidth]{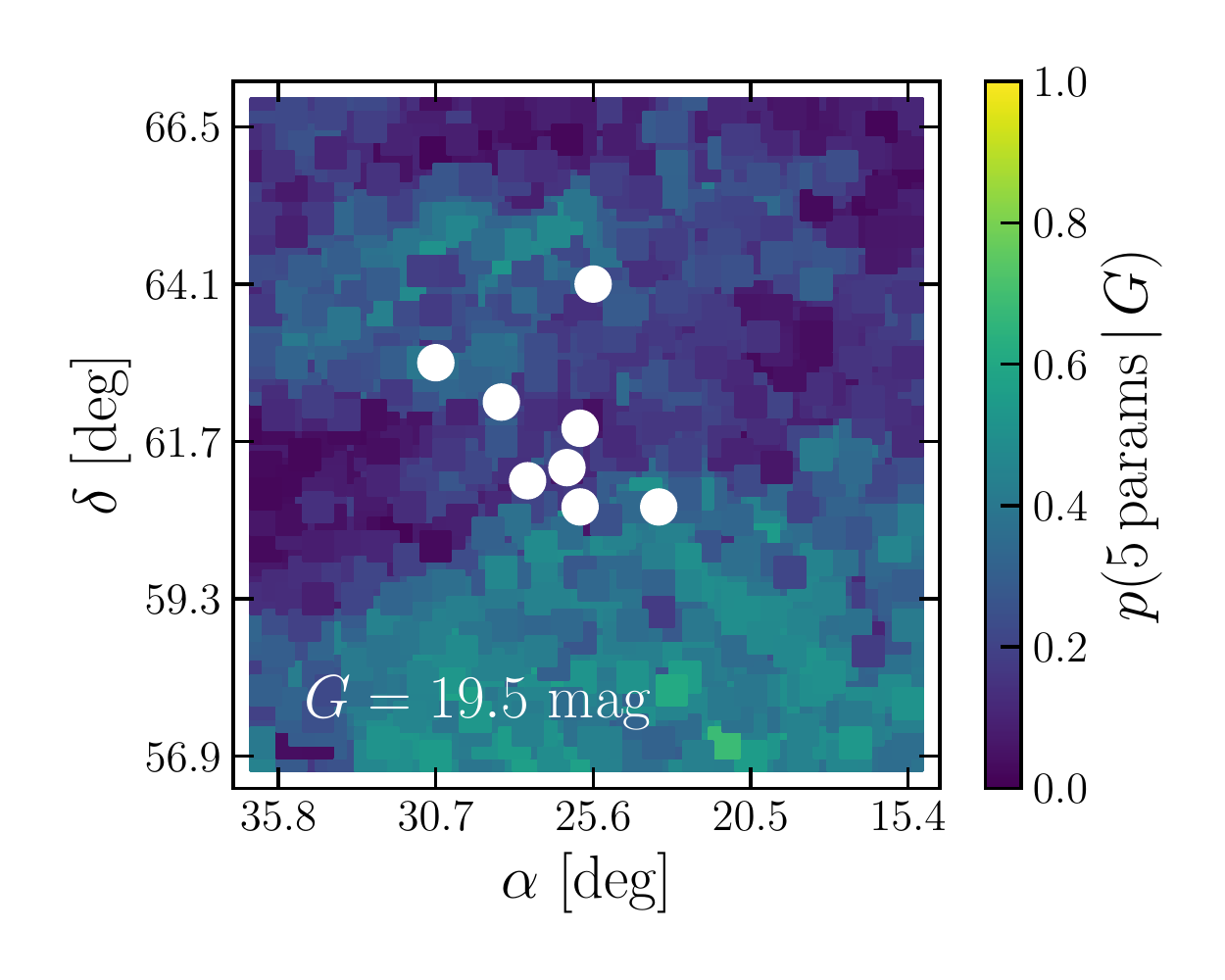}
    \end{minipage}
    \caption{Catalog's completeness in the 5 parameters solution estimated from star count ratios ($p({\rm 5 params}|G)$ term in Equation \ref{eq:completeness-5params}) for $G=18.5$ mag (left panel) and $G=19.5$ mag (right panel). White points mark the positions of star clusters. \label{fig:completeness-5params}}
\end{figure*}
In Figure \ref{fig:completeness-5params} we show as examples the two-dimensional $p({\rm 5 params}\,|\,G)$ maps computed for $G=18.5$ mag (left panel) and $G=19.5$ mag (right panel). 

The comparison between Figures \ref{fig:completeness-Gband} and \ref{fig:completeness-5params} clearly shows that the selection of sources with 5 parameters solution has a dominant impact on the final completeness. Indeed, $p({\rm 5 params}\,|\,G)$ significantly drops for $G>18.5$ mag, while $p(G)$ remains almost equal to 1 down to $G=19.5$ mag.
Hence in the following we assumed $p(G)=1$ thus simplifying Equations \ref{eq:completeness} and \ref{eq:completeness-5params} into
\begin{equation}
    p({\rm 5 params}\,\cdot\,G) \simeq \frac{k+1}{n+2} \quad \forall\,G<19.5\,{\rm mag}\,.
    \label{eq:completeness_final_5prms}
\end{equation}
In the subsequent analyses, we correct stellar counts for incompleteness according to Equation \ref{eq:completeness_final_5prms}.

\section{Physical properties of the observed area}\label{sec:coherence_physical_prop}

In Section \ref{sec:clustering_analysis} we identified a region encompassing nine star clusters embedded in 
a low-density and diffuse stellar halo (see Figure \ref{fig:fov}) lying within strict ranges in 2D velocity, position, and parallax by construction. 

In this Section, we characterize the physical properties of the area based on the Gaia photometry and the spectroscopic data.

\subsection{Differential reddening} \label{sec:differential_redd}
\begin{figure}[h!]
\centering
\includegraphics[width=0.5\textwidth]{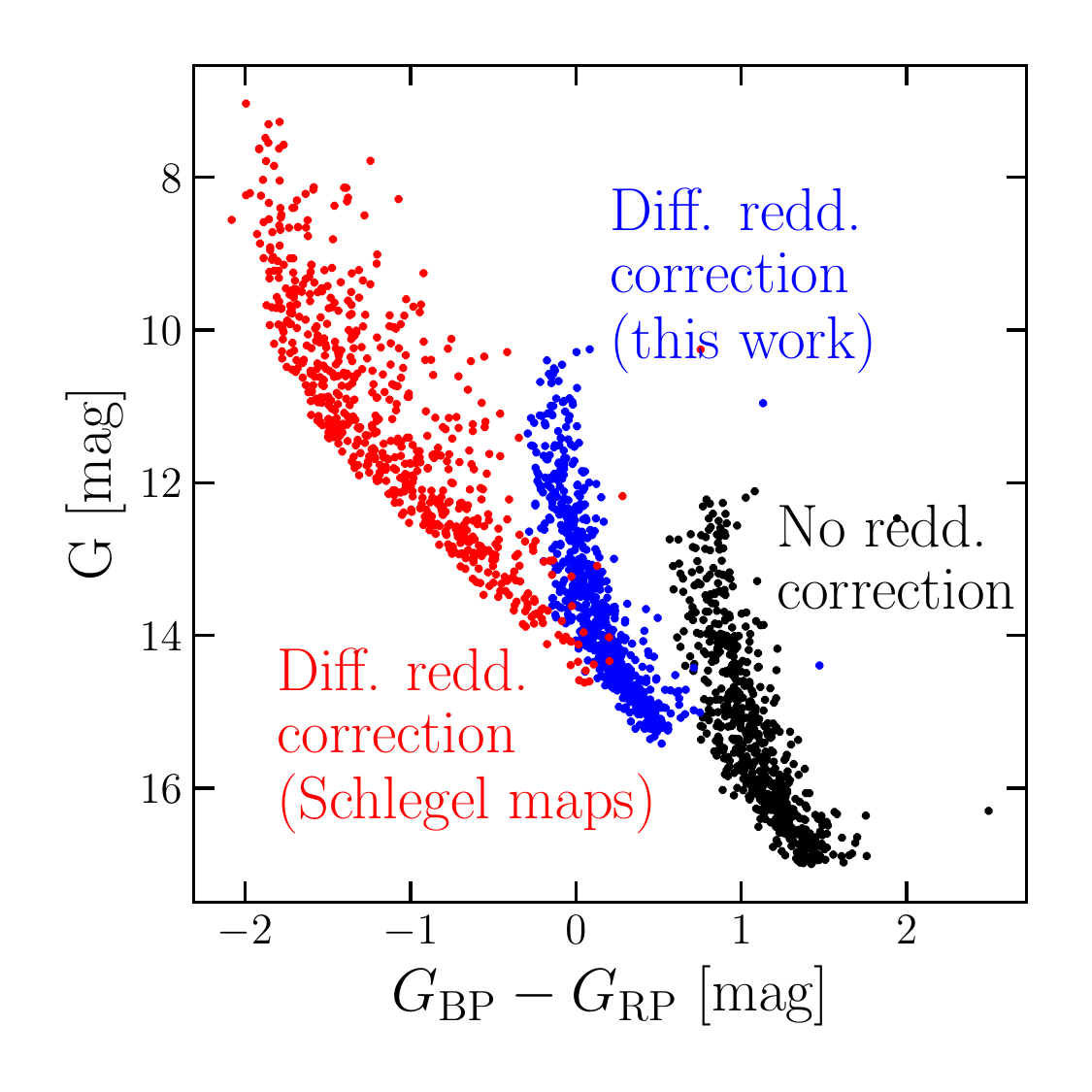}
    \caption{Color-magnitude diagram for NGC~663 members. In black the observed photometry, while in red and blue are shown the distributions of stars after correcting for differential reddening. For the former, we use Galactic extinction maps by \citet{schlegel+1998,schlafly+2011}, whereas the latter uses reddening corrections estimated in this work. \label{fig:comparison_schlegel}}
\end{figure}
\begin{figure}[h!]
    \centering
    \includegraphics[width=0.5\textwidth]{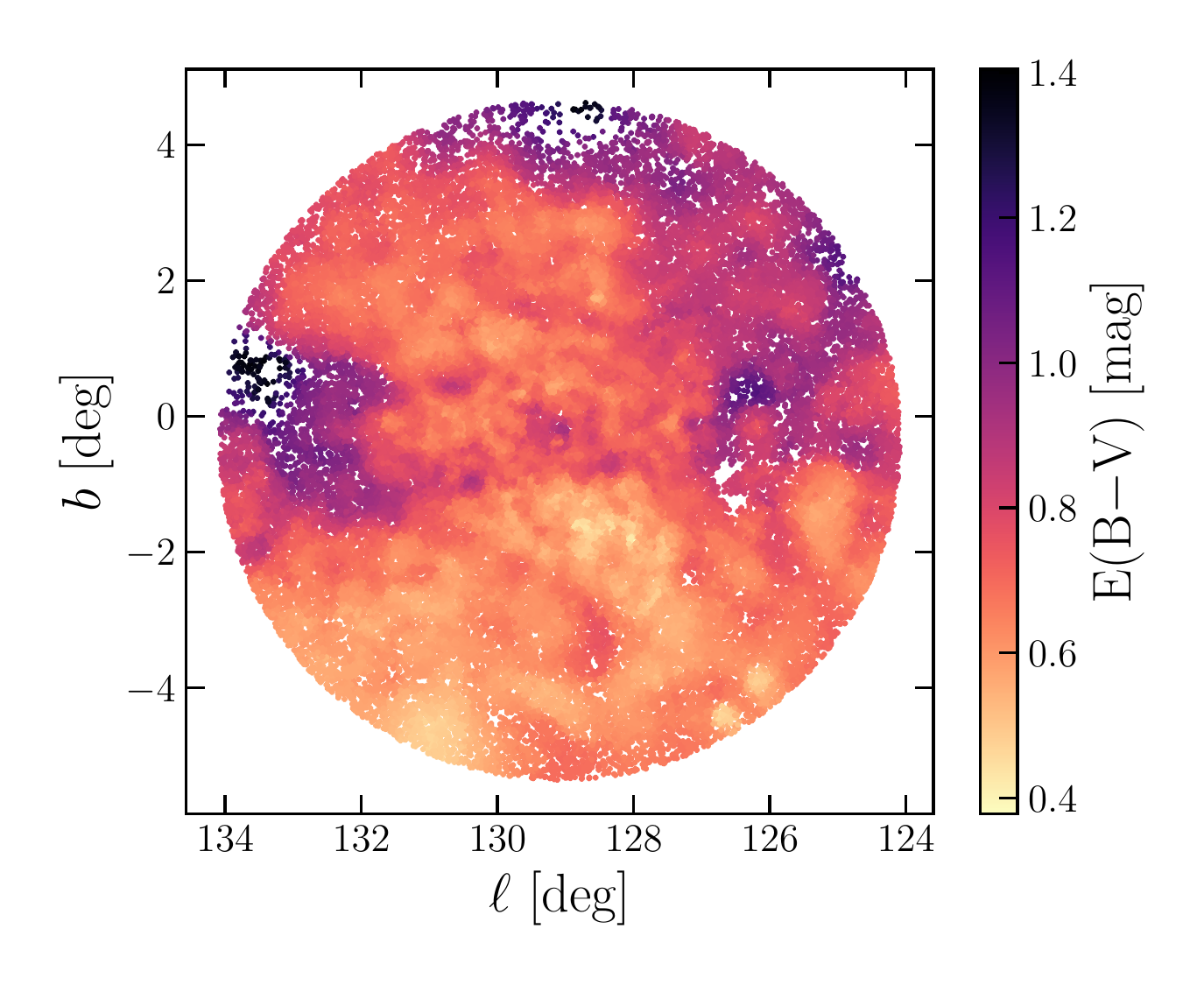}
    \caption{Two-dimensional reddening map in Galactic coordinates. The extinction has been computed star by star from color-color diagrams (see the text for further details). White areas correspond to regions devoided of stars. \label{fig:reddening}}
\end{figure}

Available Galactic extinction maps (e.g. \citealt{schlegel+1998} and recalculations from \citealt{schlafly+2011}) report
a quite significant and strongly variable (E(B$-$V)$\sim 0.5-3.5$ mag) extinction along the line of sight for the region under investigation.
Here we provide an independent estimate of the differential reddening based on a suitable color-color diagram and following the approach adopted by \citet{Dalessandro+2018}.
In particular, combining the Gaia $G$ band with the $r$, $i$, $z$ photometric bands from the Panoramic Survey Telescope and Rapid Response System \citep[Pan-STARRS data release 2,][]{panstarrs}, we constructed the ($G-r$) vs ($i-z$) color-color diagram.
Such diagram turned out to be the most suitable choice as the evolutionary sequences run almost orthogonally to the reddening vector in these colors.

We derived differential extinction star by star by minimizing differences along the reddening vector with respect to a reference system. As a reference, we chose the median color-color distribution of likely main-sequence stars (G\,$>12$ mag or G$_{\rm BP}-$G$_{\rm RP}<0.5$ mag) belonging to the cluster NGC~581.
NGC 581 stars are distributed on average at bluer colors than other stars in the field, thus suggesting they are located in a region with relatively small extinction (color excess for these stars has been derived from \citealt{schlegel+1998,schlafly+2011}: E(B$-$V)$_{\rm NGC581} \simeq 0.54$ mag). 
Afterward, for each star, we obtained the median colors of the closest 50 likely main sequence (G\,$>12$ mag or G$_{\rm BP}-$G$_{\rm RP}<0.5$ mag) neighbor stars and we determined the distance of such median value to the reference point along the reddening vector \citep[using coefficients from][]{cardelli+1989}.  
The extinction value corresponding to the derived distance is then assigned to the specific star.
To all the sources which do not fulfill the criteria of being \emph{likely main sequence stars}
and to those that do not have a counterpart in the PanSTARRS catalog, we assign the median reddening of the closest 50 neighbors.

As a representative example, we show in Figure \ref{fig:comparison_schlegel} the observed CMD of NGC~663 members as well as the ones obtained using extinction values from \citet[][in red]{schlafly+2011} and obtained in this work (in blue). 
The differential reddening corrections derived in this work nicely squeeze the sequence in the CMD compared to the observed one, thus confirming the robustness of our estimates.
On the contrary, those from \citet{schlafly+2011} significantly spread the sequence and move stars at non-physical colors, reaching $(G_{\rm BP}-G_{\rm RP})_0 = -2$ mag, thus
suggesting that the adopted values for the E(B$-$V) variations are likely overestimated.
\begin{figure*}[h!]
    \centering
    \includegraphics[width=\textwidth]{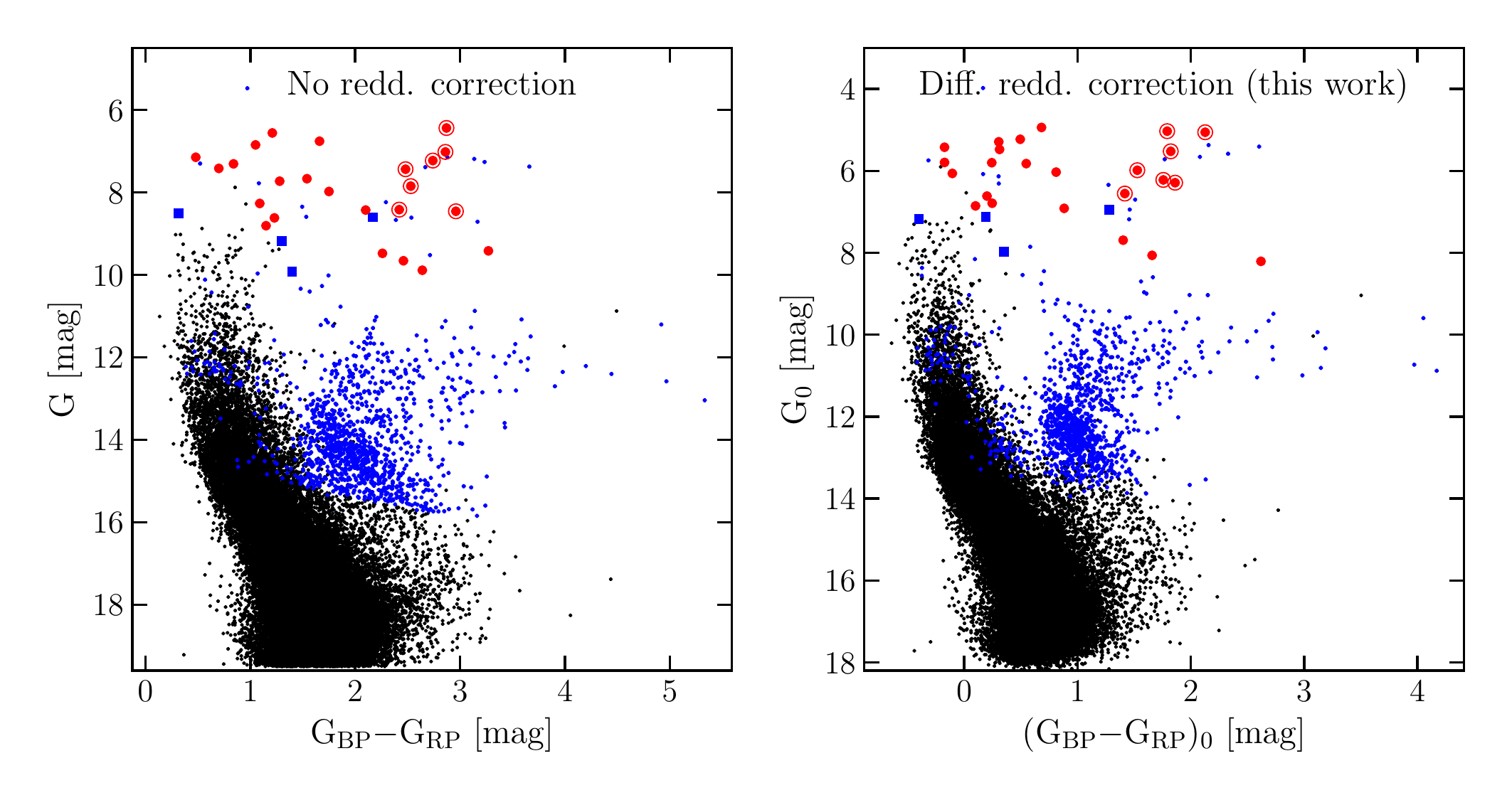}
    \caption{Observed (left panel) and differential reddening corrected (right panel) CMD for the full catalog. In blue we show stars with LOS velocity measurements from Gaia DR3 (big squares for stars considered in the LOS analysis), whereas in red are stars targeted by high-resolution spectroscopy. The subsample with chemical abundances is circled in red. \label{fig:CMD_fullcatalog}}
\end{figure*}
Finally, in Figure \ref{fig:reddening} we show the resulting reddening map as derived in this work, in which each star in the catalog is color-coded according to the inferred extinction. 
We note here that while differential reddening might play a role in shaping specific features of the 
iso-density contours shown in Figure~2 (i.e. missing sources due to locally higher extinction artificially produces underdense regions), it 
is unlikely that it impacts the overall observed density gradient across the field of view, as low-extinction regions, such as $b\lesssim-3^\circ$ (see Figure~\ref{fig:reddening}), still result under dense compared to the diffuse halo.
In Figure \ref{fig:CMD_fullcatalog} we show 
both the observed (left panel) and differential reddening corrected (right panel) CMDs of the full catalog for comparison. 

\subsection{Clusters' and halo's ages \label{sec:age_estimates}}
Age determination of young ($<100$ Myr), sparsely populated star clusters is certainly a challenge.
In fact, at these ages, the color-magnitude distribution of turn-off stars is strongly affected by stellar rotation \citep{li+2019}.
Moreover, the low number of stars and short evolutionary time scales of massive stars ($>10$ M$_\odot$) might hamper a detailed estimate of the bright and blue main sequence termination, which in turn would bias the age inferred by standard methods such as isochrone fitting. 
Notwithstanding these limitations, here we attempt to tackle this issue by 
adopting a specific approach that is only marginally sensitive to stellar rotation and minimizes the impact of low-number statistics.
In particular, we used 
a set of synthetic simple stellar populations obtained from the PARSEC database \citep{bressan2012_parsec} with $[{\rm Fe}/{\rm H}]\simeq (-0.30 \pm 0.01)$ dex
\citep[corresponding to the mean metallicity of the area][]{fanelli_etal2022}
and sampling the age range 1$-$100 Myr with a regular step of 1 Myr. We compared them with the observed cumulative luminosity functions
(CLF) in the $G$-band, after correcting for differential reddening\footnote{We use the subscript "$0$" for reddening-corrected magnitudes and not for absolute ones, i.e. they are not corrected for dimming due to the distance.} and completeness as described in Section \ref{sec:completeness}.

\begin{figure}[h!]
    \centering
    \includegraphics[width=0.5\textwidth]{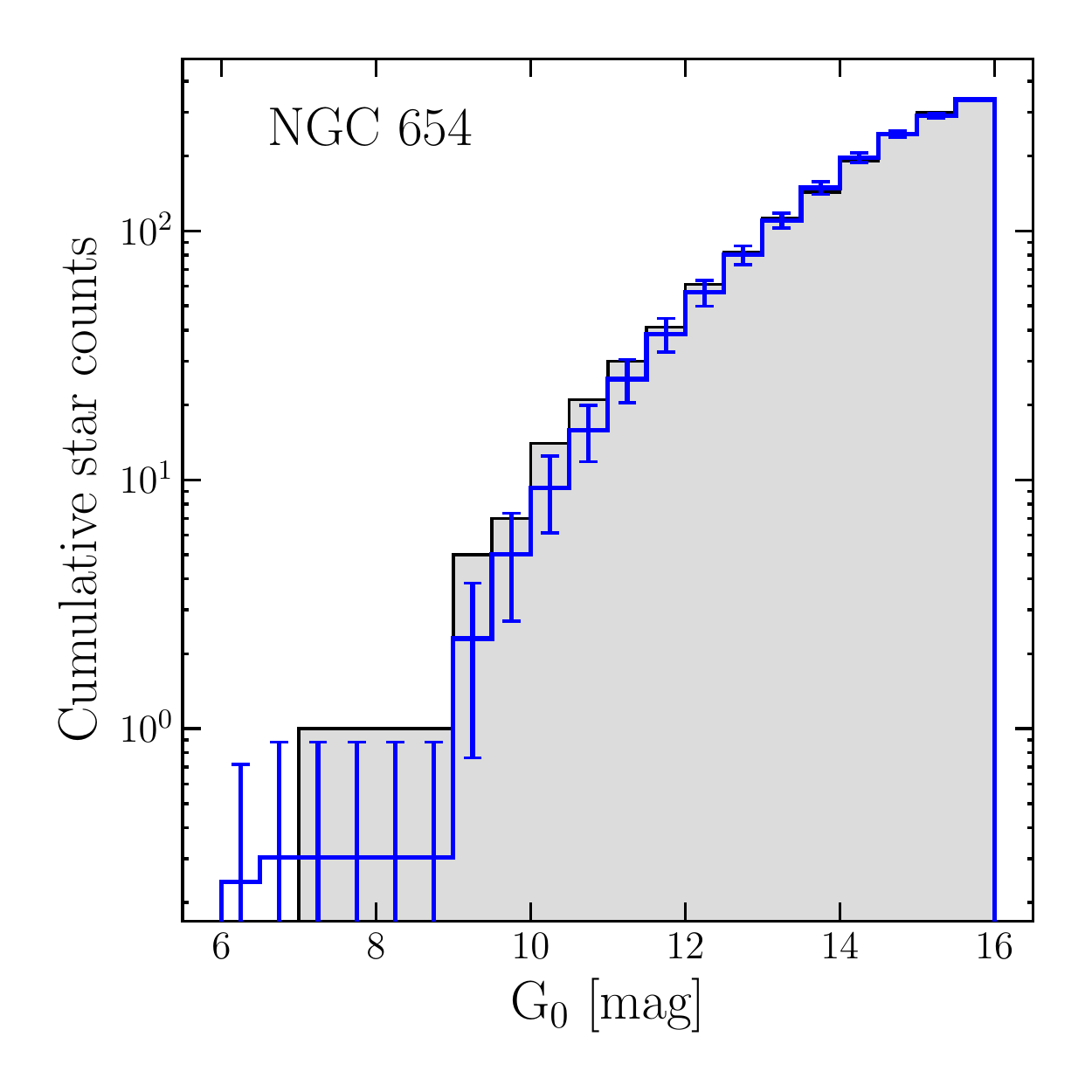}
    \caption{Cumulative luminosity function for the stellar cluster NGC~654 (gray histogram), along with the normalized histogram of a synthetic population 32 Myr old computed with the PARSEC model (blue histogram). Error bars show the standard deviation of the model's count fluctuations due to several extractions (see text for further details). Bins are $0.5$ mag wide. \label{fig:lum_func}}
\end{figure}

In order to account for number fluctuations, we randomly picked 
several times ($\simeq$100) a (virtually) independent sample of $N$ stars from each synthetic population.
The number of extracted stars $N$ has been set to be the number of objects in the synthetic population with G$_0>6$ mag after applying a normalization to the luminosity function in the range $12.5<\,$G$_0<16$ mag (at least $2$ mag fainter than the main-sequence termination for populations younger than $100$ Myr at a distance of about 3 kpc).
For each extraction, we then constructed the CLF and we determined the median CLF of all the extractions (as well as its corresponding 68\% credible region).
The median CLF obtained for different ages is then compared to the observed one (see for instance Figure \ref{fig:lum_func}) and the best fit is defined as the one that minimizes the $\chi^2$ statistics.
Comparison with the CLF has carried out up to G $=18$ mag (corresponding to about G$_0\simeq16$ mag), below which the catalog's completeness drops (see Figure \ref{fig:completeness-5params}). 

Furthermore, when comparing the synthetic CLF to the observed one, we looked for supergiant stars in the range G$_0\,<\,8$ mag and (G$_{\rm BP}-$G$_{\rm RP})_0\,>\,0.2$ mag. If present, such stars provide strong constrains on the age, hence we limited the analysis only to those ages that are able to explain the presence of evolved stars at the observed magnitudes.
This allowed us to inform the fitting procedure about the likely young age of the system even in absence of bright, blue main sequence stars.
We point out that with this procedure we assigned a narrower uniform prior to the cluster’s age.
If red supergiants were not present we did not apply any selection on the age.

\begin{figure*}[h!]
    \centering
    \includegraphics[width=\textwidth]{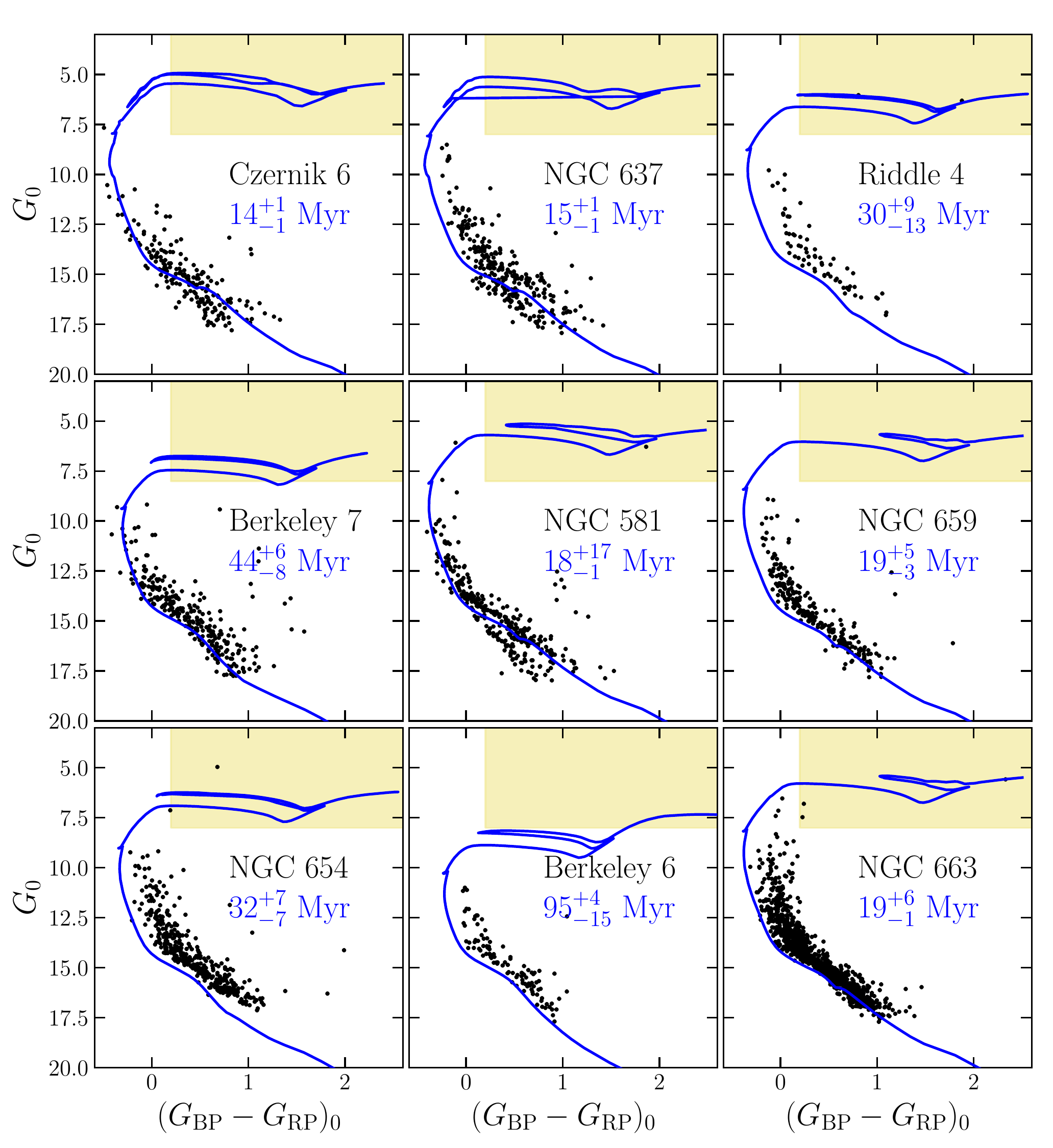}
    \caption{Color-magnitude diagrams of clusters' members (black dots) corrected for differential reddening. For each cluster, the best-fit isochrone is shown in blue and the median age along with the 68\% credible interval are also reported. The shaded areas show the region inside which we flagged stars as supergiants. \label{fig:ages}}
\end{figure*}

Color-magnitude diagrams for each cluster are shown in Figure \ref{fig:ages} along with best-fit isochrones. 
The nine clusters result to have ages in a narrow range $14-44$ Myr. The only exception is Berkeley 6 for which we derived an age of $95^{+4}_{-15}$ Myr. We further notice a slight mismatch between isochrones and data visible especially in the clusters Riddle~4 and NGC~654. Such discrepancy likely results from local underestimations of the differential reddening which in turn would bias the age inference toward older ages.

Nevertheless, typical errors in age estimates are about $10-15$ Myr (Figure \ref{fig:ages}).
They account only for uncertainties arising from the fitting procedure, albeit errors in the differential reddening, distance, and also wrong membership assignment might be important sources of uncertainties.
However, we stress that we are mainly interested in constraining relative ages rather than absolute ones.

Comparison with ages from the literature shows qualitatively overall agreement as they range between $15-38$ Myr for the clusters under study \citep{cantat-gaudin+2020}.
The only exception is Berkeley~6 for which \citet{cantat-gaudin+2020} report an age of about $200$ Myr, consistent with the system being older than other clusters. 

Finally, the same analysis has been carried out for the stellar halo,
i.e. all the stars which did not belong to any cluster according to the membership probabilities assigned by the clustering algorithm,
finding that its age ($\sim16^{+1}_{-1}$ Myr) is consistent with the ones of the clusters embedded within it.

\subsection{LOS velocity distribution and iron content} \label{sec:coherence6Dlos}
We investigated the LOS velocity and the metallicity distributions in the region 
using the TNG-GIARPS spectroscopic catalog presented in Section \ref{sec:catalog} and marked in red in Figure~\ref{fig:CMD_fullcatalog},
and we compared them with those expected for the surrounding Galactic field obtained from the Besan\c con Milky-Way model \citep{robin+2003_besanconModel} after applying the same parallax and proper motion selections.
We computed LOS velocities for 24 stars, five of which are cluster members while the remaining 19 belong to the halo. Among them, chemical abundances are available for the 7 red supergiants (double red circles in Figure~\ref{fig:CMD_fullcatalog}), one of which belongs to NGC~581.

We also note that, while Gaia DR3 \citep{gaiaDR3_2022} provides LOS velocity for 1164 selected stars, their color-magnitude distribution (shown in Figure \ref{fig:CMD_fullcatalog}) suggests they are mostly field interlopers. Nevertheless, some bright stars likely members of the system have LOS measurements from Gaia. We, therefore, selected those stars with $G_0<8$ mag (removing objects belonging to the older disk population) and with \texttt{rv\_expected\_sig\_to\_noise}$>5$ and \texttt{rv\_renormalised\_gof}$<2$ \citep[thus selecting sources with reliable LOS velocity, see][]{katz+2022_radialVelGaiaDR3}. Out of the 1164 stars, only 4 fulfilled these criteria and were thus included in the catalog (shown with larger blue square markers in Figure \ref{fig:CMD_fullcatalog}).
\begin{figure*}[h!]
    \centering
    \begin{minipage}{0.475\textwidth}
        \includegraphics[width=\textwidth]{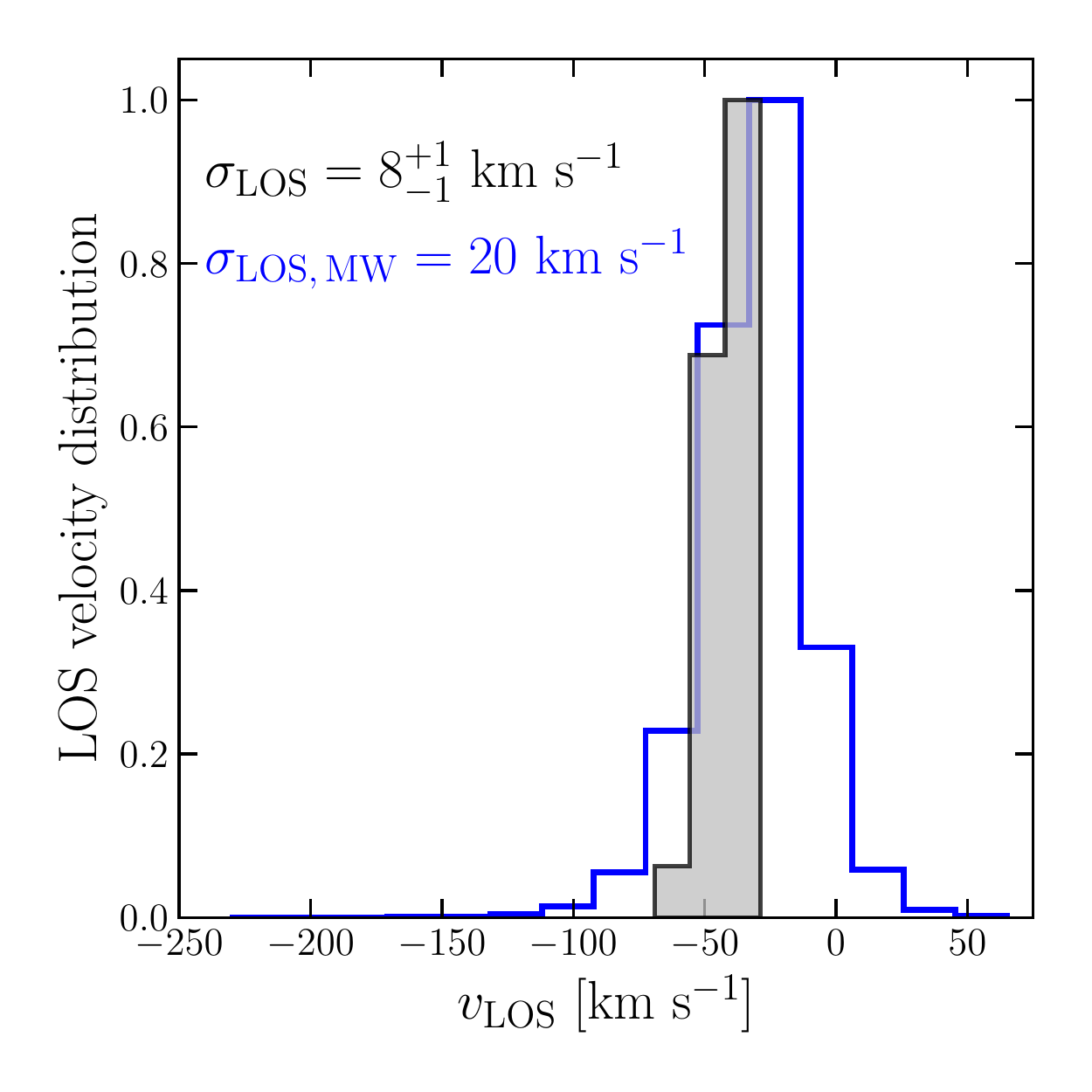}
    \end{minipage}
    \begin{minipage}{0.475\textwidth}
        \includegraphics[width=\textwidth]{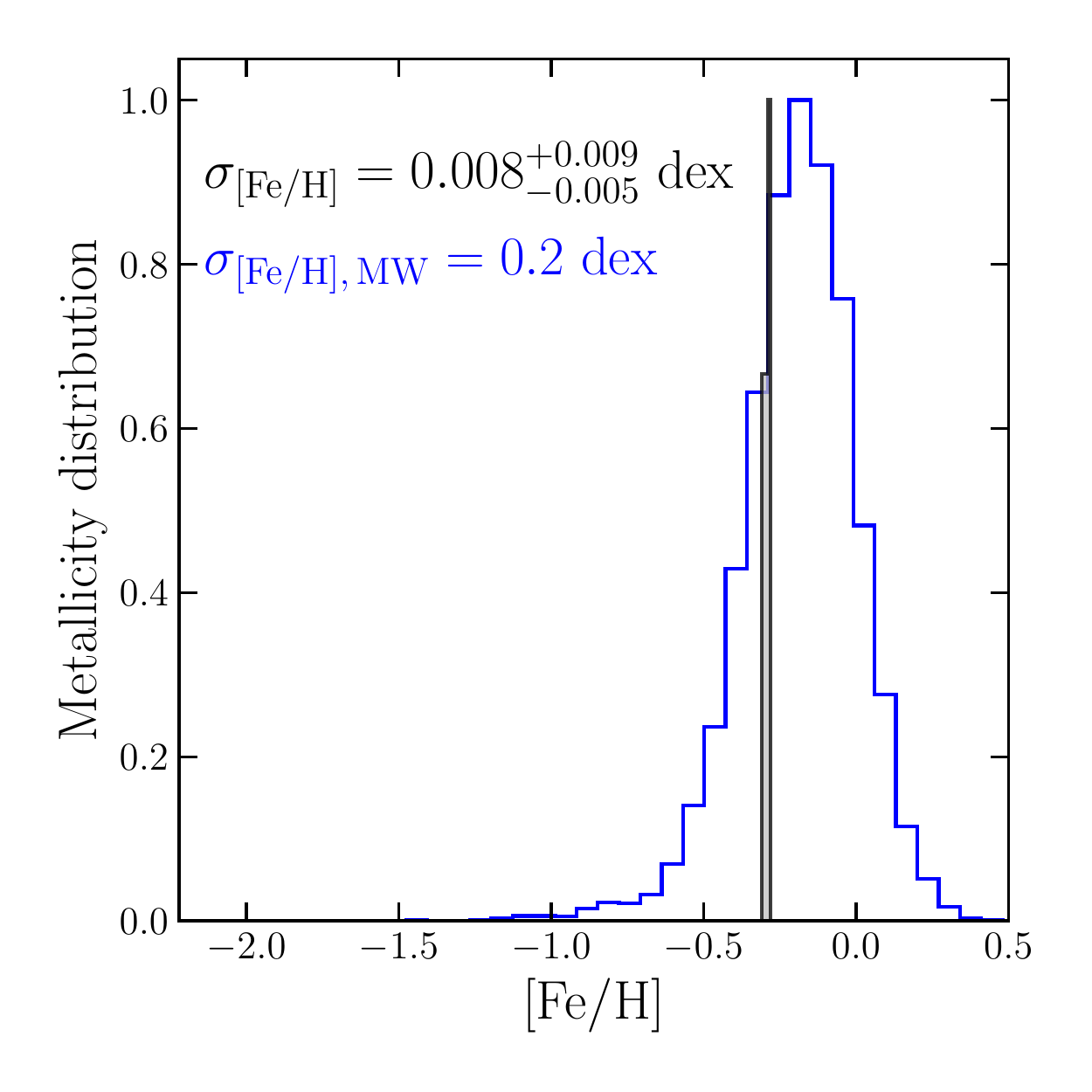}
    \end{minipage}
    \caption{ 
    Line-of-sight velocity (left panel) and iron-over-hydrogen abundance (right panel) distributions for members of the selected structure (gray histogram) and for a Milky-Way model (blue histogram). The intrinsic dispersions are also shown in the top-left corners. 
    All the distributions have been scaled by a constant factor for visualization purposes only.  \label{fig:comparisonLOSdistr_lisca-mw}}
\end{figure*}

In Figure \ref{fig:comparisonLOSdistr_lisca-mw} we show the distributions in LOS velocity (left panel, constructed with 24 stars from the high-resolution spectroscopic catalog plus 4 stars from Gaia DR3) and in $[{\rm Fe}/{\rm H}]$ abundance (right panel, for the seven red supergiants), superimposed to the distributions of the surrounding Galactic field. 
The intrinsic widths of the observed distributions were inferred by means of a maximum likelihood approach (accounting for individual errors on measurements) and in Figure \ref{fig:comparisonLOSdistr_lisca-mw} we report their median values along with the 68\% credible intervals.
In particular, we obtained a LOS velocity dispersion $\sigma_{\rm LOS}=8^{+1}_{-1}$ km s$^{-1}$ (to be compared with $\sigma_{\rm LOS, MW} = 20$ km s$^{-1}$) and a mean velocity $\langle v_{\rm LOS} \rangle=-41^{+2}_{-2}$ km s$^{-1}$, whereas for the metallicity we obtained a dispersion $\sigma_{[{\rm Fe}/{\rm H}]}=0.008^{+0.009}_{-0.005}$ dex (opposed to $\sigma_{[{\rm Fe}/{\rm H}]}=0.2$ dex for the Galactic field) and a mean metallicity $\langle [{\rm Fe}/{\rm H}]\rangle=-0.30^{+0.01}_{-0.01}$ dex.

Interestingly, the observed distributions of stars in the region are significantly narrower than the ones expected for a randomly selected group of co-moving Galactic stars, thus strengthening the evidence of kinematic coherence and suggesting a significant chemical homogeneity. 
Also, the consistency in both LOS velocity and chemical content between cluster and halo stars further validates the assumption of a physical and coherent structure embedding the star clusters.

In addition, literature data about bulk clusters' LOS velocity support the kinematic coherence of all clusters but one. In fact, 
the LOS velocity reported for Berkeley 6 ($v_{\rm LOS} \simeq (-89 \pm 52)$ km s$^{-1}$) \citep{tarricq+2021} is significantly lower than the system's bulk velocity. However, we note that this value is still compatible within the huge uncertainty as only two stars have been used for its estimate. 
Moreover, \citet{spina+2021} measured the iron content for one member of Berkeley 6 to be around $[{\rm Fe}/{\rm H}]\sim -0.179$ dex significantly higher than $[{\rm Fe}/{\rm H}]\simeq-0.3$ dex although we stress again that chemical abundances have been derived only for one cluster's member whose membership probability is $<30\%$ \citep{spina+2021}.
Therefore, better constraints on the stellar membership, age, and three-dimensional velocity are needed before drawing any conclusion about the role of Berkeley~6 in the system.

\section{Structural and kinematic properties of the diffuse stellar halo} \label{sec:kin_struct}
\subsection{Density distribution}
We constructed the number density profile of the diffuse stellar halo.
We took as the system's center the center of mass of stars with $M\geq2$ M$_\odot$.
Firstly, celestial coordinates have been converted into local Cartesian ones, assuming the centroid of the system as an initial guess for the system's center.
After that, the center of mass has been computed in Cartesian coordinates and it has been converted back into celestial ones, obtaining $(\alpha_{\rm CM};\delta_{\rm CM} ) = (26.4559;\,61.7865)$ degree.

We binned stars radially with respect to this center and we set the width of each radial annulus to contain 2500 sources each.
Radial shells were then split into four angular sectors where the density has been computed simply as the ratio between the number of stars and the sector's area. The final shell density and error were the mean and standard deviation of the four measurements respectively. Finally, we also accounted for Poissonian error in each bin by summing in quadrature to the standard deviation a term $1/\sqrt{N_{\rm shell}}$, with $N_{\rm shell}$ being the number of stars within the shell. 

\begin{figure}[h!]
\centering
\includegraphics[width=0.5\textwidth]{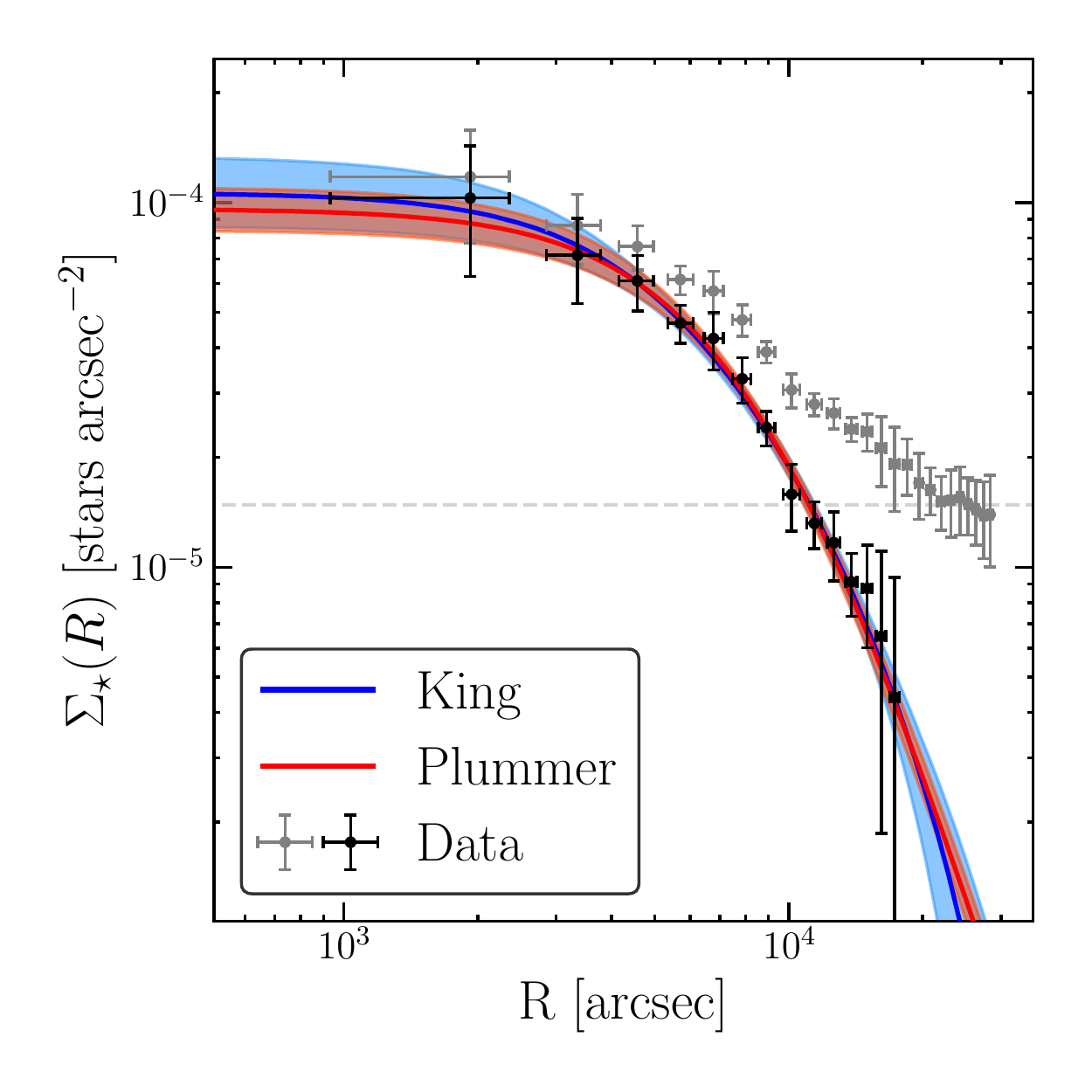}
\caption{
Stellar number density profile of an 8$^\circ$-wide region around the system's center of mass and considering stars brighter than $G=18$ mag. The observed profile is reported in gray, while the \emph{intrinsic one} (after background subtraction) is shown in black. Plummer (in red) and King (in blue) models are also shown, along with the corresponding 68\% credible regions constructed from the posterior samples.
\label{fig:fit_profile_halo}}
\end{figure}

In Figure \ref{fig:fit_profile_halo}, we show the number density profile for sources out to 8$^\circ$ from the system's center of mass and with $G\leq18$ mag. 
When studying the density distribution, we temporarily extended the catalog up to 8$^\circ$ from the system's center in order to assess the background density, while
the latter selection in $G$ was a good compromise between the catalog's completeness and statistics. 
Interestingly, the observed density resembled a cluster-like profile over about a factor of 10 in density.
At about $R\gtrsim6^\circ$, the density profile flattens, and we estimated the \emph{background density} as the weighted mean of bins 
at distances larger than 6$^\circ$ from the adopted center, obtaining $\Sigma_{\star,\rm background} \simeq 1.5 \times10^{-5}$ stars arcsec$^{-2}$, that is then subtracted to the observed profile.

Finally, we fitted the density distribution within 5$^\circ$ using King \citep{king1962} and Plummer \citep{Plummer1911} models, which are typically adopted to reproduce stellar clusters' density profiles. All the free parameters were constrained assuming a $\chi^2$ likelihood, uniform priors (in logarithm), and exploring the parameters' space with a Markov Chain Monte Carlo (MCMC) technique using the Python package \texttt{emcee}\footnote{\url{https://emcee.readthedocs.io/en/stable/}.} \citep{foreman-mackey_emcee}. 

In Figure \ref{fig:fit_profile_halo} we, therefore, show the density profile (before, in gray, and after, in black, the background subtraction) along with the best-fit models and the associated errors. Both models provided a nice description of the data.

\subsection{Kinematic properties}
We investigated the kinematic properties of the stellar halo by further selecting stars fulfilling the following astrometric quality selection criteria \citep[][]{Lingedren+2021}: \texttt{ruwe} $\leq1.4$, \texttt{astrometric\_gof\_al} $\leq1$ and \texttt{astrometric\_excess\_noise} $\leq1$ mas (if \texttt{astrometric\_excess\_noise\_sig} $>2$), thus excluding those sources for which the standard five-parameter solution does not provide a reliable fit of the observed data.

Firstly, we accounted for perspective effects induced by the system's bulk motion \citep{vanLeeuwen2009} on the $\mu_{\alpha *}$ and $\mu_\delta$ components: we thus corrected the velocities for each star assuming a bulk average motion of $(\overline{\mu_{\alpha *}}; \overline{\mu_\delta}) = (-1.14; -0.33)$ mas yr$^{-1}$ (estimated using Gaia data for sources with reliable astrometry) and the mean LOS velocity obtained from the spectroscopic catalog supplemented with Gaia DR3 data (see Section \ref{sec:coherence6Dlos}).
Owing to the large area of the sky covered by our data, the magnitude of the perspective correction resulted as nonnegligible, reaching up to $0.2$ mas yr$^{-1}$ (about $2.8$ km s$^{-1}$) at $5^\circ$ from the center, hence caution must be taken when interpreting results at such large angular scales.

Looking at the distribution of velocities onto the plane of the sky may offer a first glimpse of the dynamic state of the system. 
We thus performed a centroidal Voronoi tessellation \citep{cappellari+2003}, exploiting the density profile shown in Figure \ref{fig:fit_profile_halo} such that each bin contains about the same number of stars.
\begin{figure}[h!]
\centering
    \includegraphics[width=0.5\textwidth]{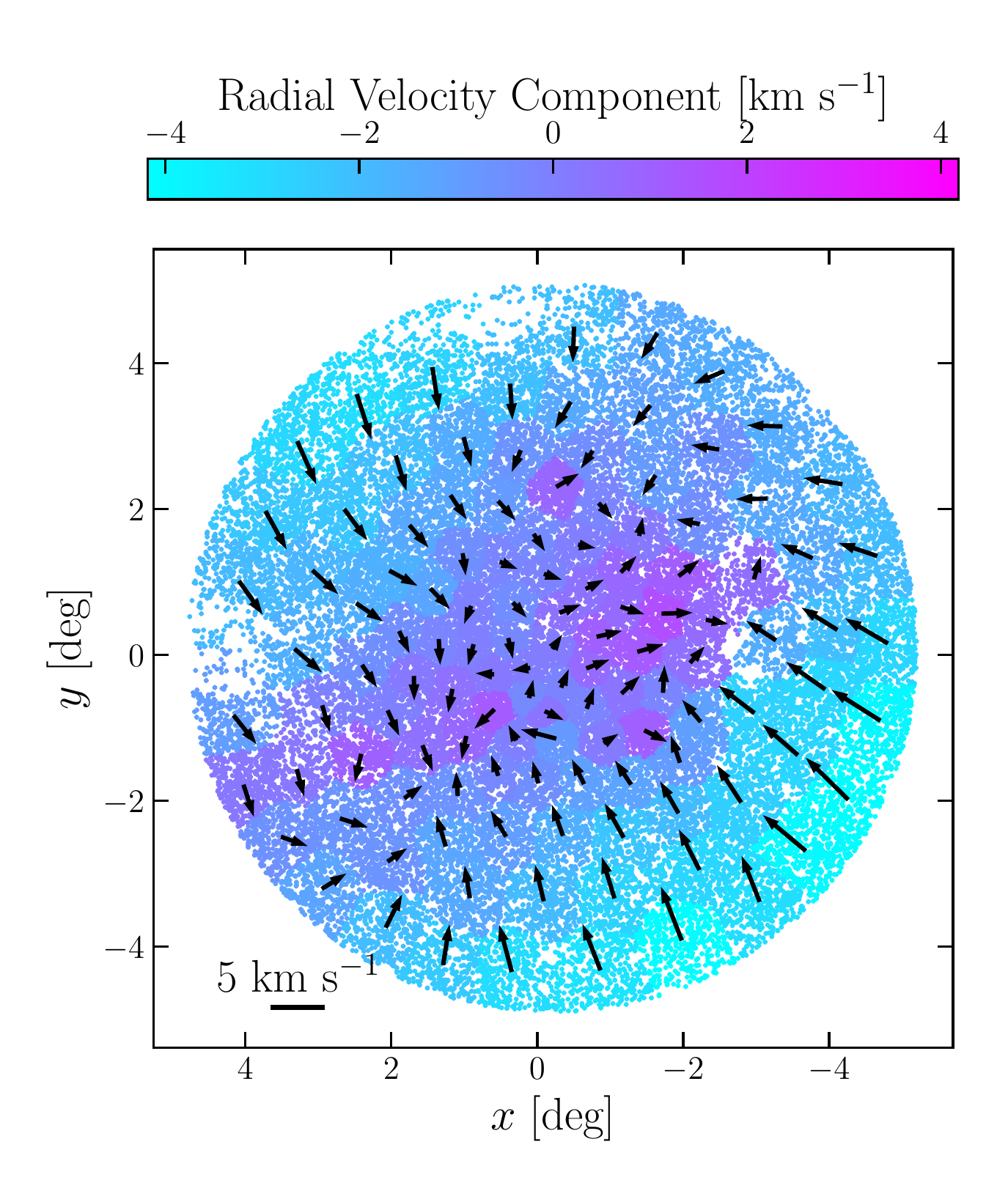}
    \caption{Two dimensional distribution of stars in locally Cartesian coordinates $(x;y)$. Stars are color-coded according to the inferred mean radial component of the velocity in their Voronoi bin, while black arrows show the mean velocity vector in each bin. 
    Radial velocity is defined as positive if pointing away from the center, thus $\overline{v_{\rm R}}>0$ means expansion and vice versa.
    The velocity scale is shown in the bottom left corner. \label{fig:stream_voronoi}}
\end{figure}

Radial velocities\footnote{Throughout the paper we shall refer as \emph{radial velocity} to the radial component of the velocity projected onto the plane of the sky, whereas as line-of-sight velocity to the component aligned with the observer's line-of-sight.} in each bin have been inferred by means of an MCMC exploration assuming as likelihood \citep[see e.g.][]{pryor_meylan_1993,raso+20}
\begin{eqnarray}
    \ln\mathcal{L} = -\frac{1}{2}\sum_{i}\Bigg[ \frac{(v_{\rm R,i} - \overline{v_{\rm R}})^2}{\sigma_{\rm R}^2 + e_{\rm R,i}^2} + \ln(\sigma_{\rm R}^2 + e_{\rm R,i}^2) \nonumber \\
    +\,\frac{(v_{\rm T,i} - \overline{v_{\rm T}})^2}{\sigma_{\rm T}^2 + e_{\rm T,i}^2} + \ln(\sigma_{\rm T}^2 + e_{\rm T,i}^2)\Bigg]
    \label{eq:likelihood_meandisp}
\end{eqnarray}
where $v_{\rm X,i}$ and $e_{\rm X,i}$ with ${\rm X}\in\{{\rm R, T}\}$ are the radial (R) and tangential (T) components of the velocity and error for the $i$-th star respectively.
Furthermore, we assumed uniform priors in the logarithms of the velocity dispersion ($\sigma_{\rm R}$ and $\sigma_{\rm T}$) and uniform priors in the mean velocities ($\overline{v_{\rm R}}$ and $\overline{v_{\rm T}}$).
In Figure \ref{fig:stream_voronoi} we show the mean velocity vectors for each tile and we color-coded stars in each Voronoi bin according to the mean radial velocity inferred in the bin.
The arrows' directions clearly show a contraction of the external regions (reaching speeds up to $\sim 4-5$ km s$^{-1}$) also confirmed by the color distribution of stars in Figure \ref{fig:stream_voronoi}.
Interestingly, 
in the central regions (R$<1-2$ deg), a mild expansion of the order of $\simeq1$ km s$^{-1}$ is observed, mainly visible in the purplish bins.

\begin{figure}[h!]
\centering
    \includegraphics[width=0.5\textwidth]{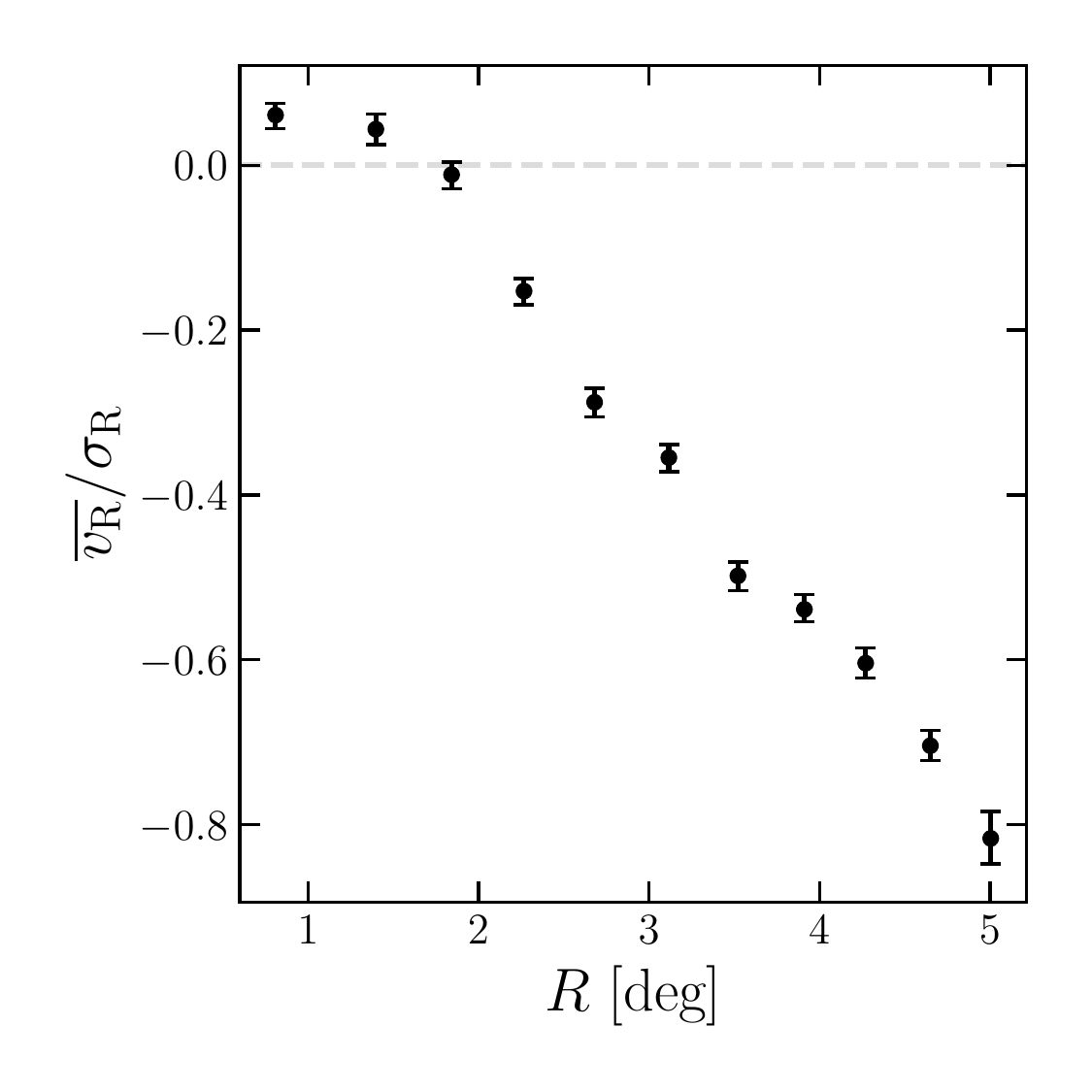}
    \caption{Radial profile of the ratio between the radial mean velocity and the radial velocity dispersion computed in spherical shells. Black dots show the median values while quoted errors are the 16th and 84th percentiles of the distributions. The dashed horizontal line shows the zero expansion/contraction level.  \label{fig:radial_contraction_profiles}}
\end{figure}

The same pattern emerges when computing the mean radial velocity $\overline{v_{\rm R}}$ in spherical shells (see Equation \ref{eq:likelihood_meandisp}), as shown in Figure \ref{fig:radial_contraction_profiles}. 
In particular, the innermost 2$^\circ$ show a flat, slightly positive profile ($\overline{v_{\rm R}}/\sigma_{\rm R}>0$) indicating central expansion, although the radial motion is highly dominated by \emph{random} motion ($\overline{v_{\rm R}} / \sigma_{\rm R} < 0.1$). 
However, moving toward larger radii, the contraction ($\overline{v_{\rm R}}/\sigma_{\rm R}<0$) becomes increasingly more prominent and the radial motion more \emph{ordered}.

The robustness of the contraction pattern (Figures \ref{fig:stream_voronoi} and \ref{fig:radial_contraction_profiles}) was tested against the assumption of a particular bulk line-of-sight velocity when accounting for perspective expansion (since we had only a few measurements). We considered the worst-case scenario in which stars followed the LOS velocity distribution expected for the MW field stars in the region ($\overline{v_{\rm LOS}} \simeq -35$ km s$^{-1}$ and $\sigma_{\rm LOS, MW} \simeq 20$ km s$^{-1}$, see Figure \ref{fig:comparisonLOSdistr_lisca-mw}).

The transverse motions were then corrected by randomly assigning to each star a LOS velocity extracted from the MW-like distribution.
We iterated such procedure 100 times finding that the velocity pattern observed was weakly affected by our mean bulk motion assumption and the contraction showed in Figure~\ref{fig:radial_contraction_profiles} was always recovered.

\subsection{Mass dependence analysis} \label{sec:mass_analysis}
\begin{figure}[h!]
    \centering
    \includegraphics[width=0.5\textwidth]{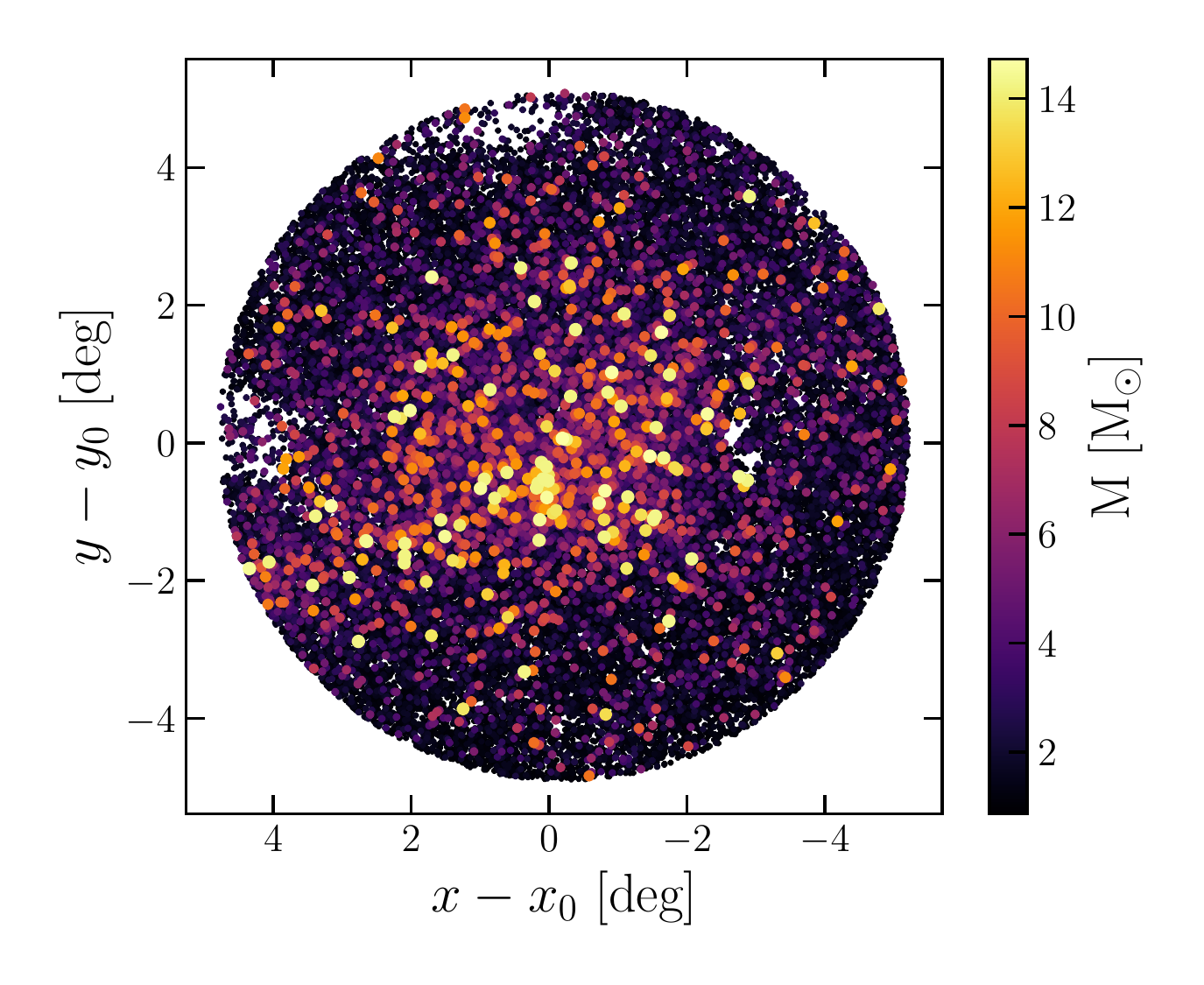}
    \caption{Spatial distribution of stars color-coded by mass.
    Masses have been obtained by mass-absolute magnitude relation (see text for further details). \label{fig:masses}}
\end{figure}
\begin{figure}[h!]
    \centering
    \includegraphics[width=0.5\textwidth]{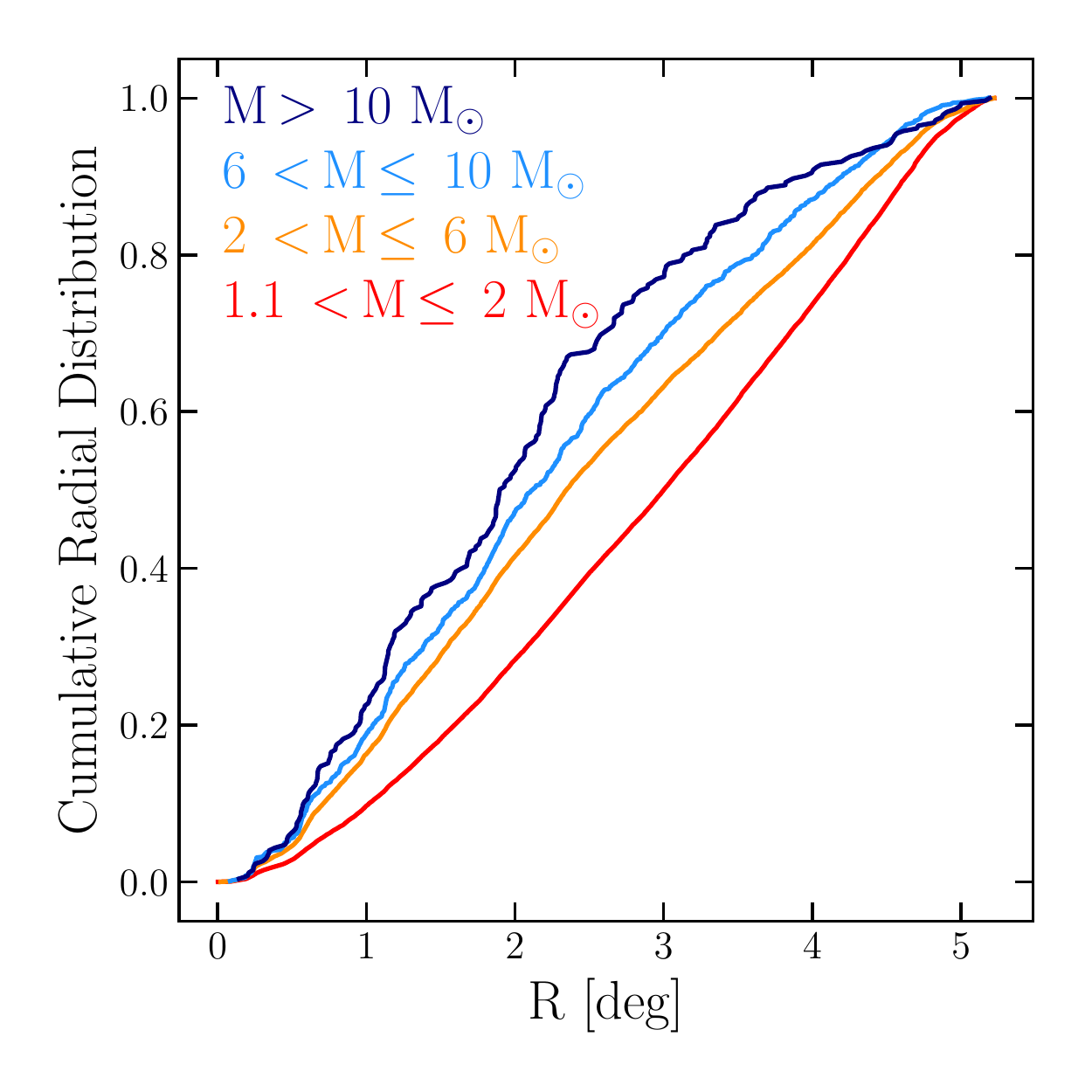}
    \caption{Cumulative radial profiles constructed in four different mass bins from stars with masses between 1.1$-$2 M$_\odot$ (in red) up to stars more massive than 10 M$_\odot$ (in dark blue). Stellar counts have been corrected for incompleteness star-by-star. \label{fig:cumulative_profiles}}
\end{figure}
\begin{figure*}[h!]
    \centering
    \includegraphics[width=\textwidth]{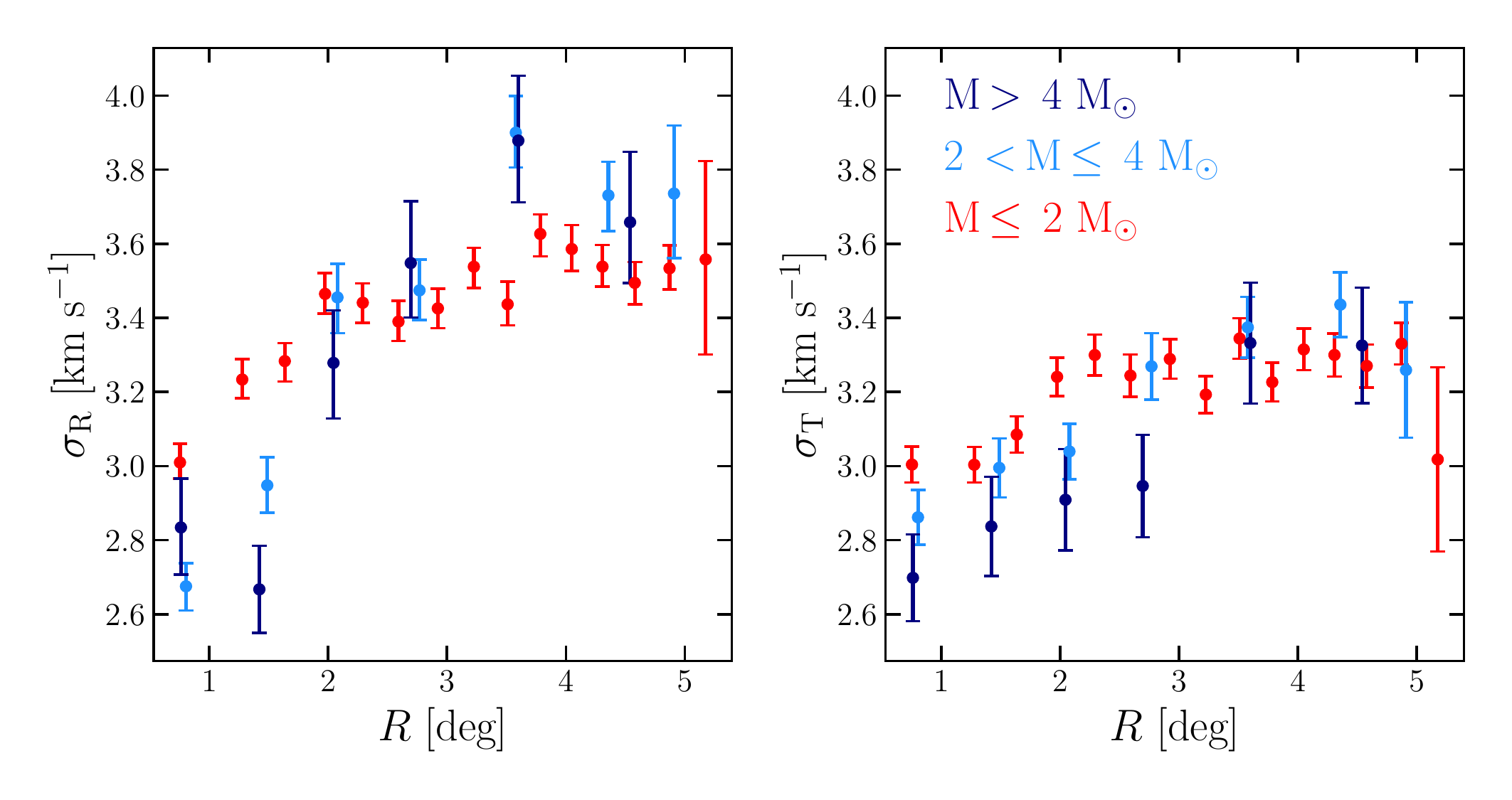}
    \caption{Velocity dispersion profiles in the radial (left panel) and tangential (right panel) components, in three different mass bins depicted with different colors: red markers for the lowest massive bin while dark-blue markers for the more massive one. \label{fig:equipartion_dispersion}}
\end{figure*}
The presence of a mass spectrum has a non-negligible role in the dynamics of both young and old stellar systems.
Old stellar clusters are indeed known to naturally develop mass segregation 
due to two-body interactions that cause significant kinetic energy exchange among stars and cause massive stars to sink towards the cluster's center \citep{binneyTremaine2008}.
However, evidence of mass segregation has also been found in younger Galactic clusters \citep[e.g.][]{Hillenbrand-Hartmann+1998,Gouliermis+2004,stolte+2006,evansOh_2022}, 
thus possibly implying a connection with the early stages of cluster formation \citep{mcmillan+2007,allison_etal2009,livernois+2021}.
In addition, numerical simulations showed that during the violent relaxation phase, young stellar systems can start developing a dependence of the kinematical properties (rotation, velocity dispersion) on the stellar mass \citep[see e.g][]{livernois+2021}.
The investigation of possible mass-dependent dynamical properties is therefore crucial to shed further light on the dynamics of young stellar systems.

We estimated stellar masses using theoretical M$-$G$_0$ relation for zero-age main sequence stars. Such relation was obtained from the PARSEC models for a population of 14 Myr old (the youngest age we estimate) with $[{\rm Fe}/{\rm H}] = -0.3$ dex.
Stellar masses have been therefore derived by interpolation of this relation, and in Figure \ref{fig:masses} we show the spatial distribution of stars color-coded by their mass. 

Figure \ref{fig:masses} indicates that massive stars (yellower markers) are more centrally concentrated than lower mass ones (darker markers): we thus quantitatively investigated this feature by looking at the cumulative profiles in different mass bins, after correcting for completeness (see Figure \ref{fig:completeness-5params}) by assigning to each star a weight $=$1/completeness. We limited the analysis to stars more massive than 1.1 M$_\odot$ to avoid very low completeness values.
Figure \ref{fig:cumulative_profiles} shows that the massive stars exhibited a more centrally concentrated spatial distribution than the lower mass ones.
In addition, a Kolmogorov-Smirnov test confirmed the statistical significance of such result ($p<0.003$) for every combination of mass bins. In Appendix \ref{appendix:completeness} we show the completeness distributions for each mass bin.

Finally, we have looked for evidence of a dependence of the kinematic properties on the stellar masses. 
In Figure \ref{fig:equipartion_dispersion} we show the velocity dispersion profiles constructed for different mass bins, specifically for stars with mass $M\leq2$ M$_\odot$ (red markers), 2~M$_\odot\,<M\leq4$ M$_\odot$ (light-blue markers) and $M>4$ M$_\odot$ (dark-blue markers).

Our analysis reveals mild evidence of a dependence of the velocity dispersion on the stellar mass within about 2$^\circ$ from the center in both the radial and tangential components: more massive stars show a slightly smaller velocity dispersion.
At radii $R>2-3$ deg the velocity dispersion profiles become indistinguishable and no signs of equipartition are found.
The trend is consistent with that expected for a stellar system that has started to evolve towards energy equipartition during its early evolutionary phases and is in general agreement with the trends found in the simulations presented in \citet{livernois+2021}. We will further discuss this point in Section \ref{sec:simulation}.

Finally, we observed an increase in the velocity dispersion moving away from the center in every mass bin (variations up to $\sim0.6$ km s$^{-1}$, see Figure \ref{fig:equipartion_dispersion}).
Several effects might be at play in driving such a pattern, for instance, deviation from spherical symmetry (see for instance the color-coded map in Figure \ref{fig:stream_voronoi}, whose expansion pattern was clearly non-spherical), a non-constant mean velocity within bins \citep[as noted by][]{daRio+2017} and possibly the presence of residual field interlopers.

\begin{figure}[h!]
    \centering
    \includegraphics[width=0.5\textwidth]{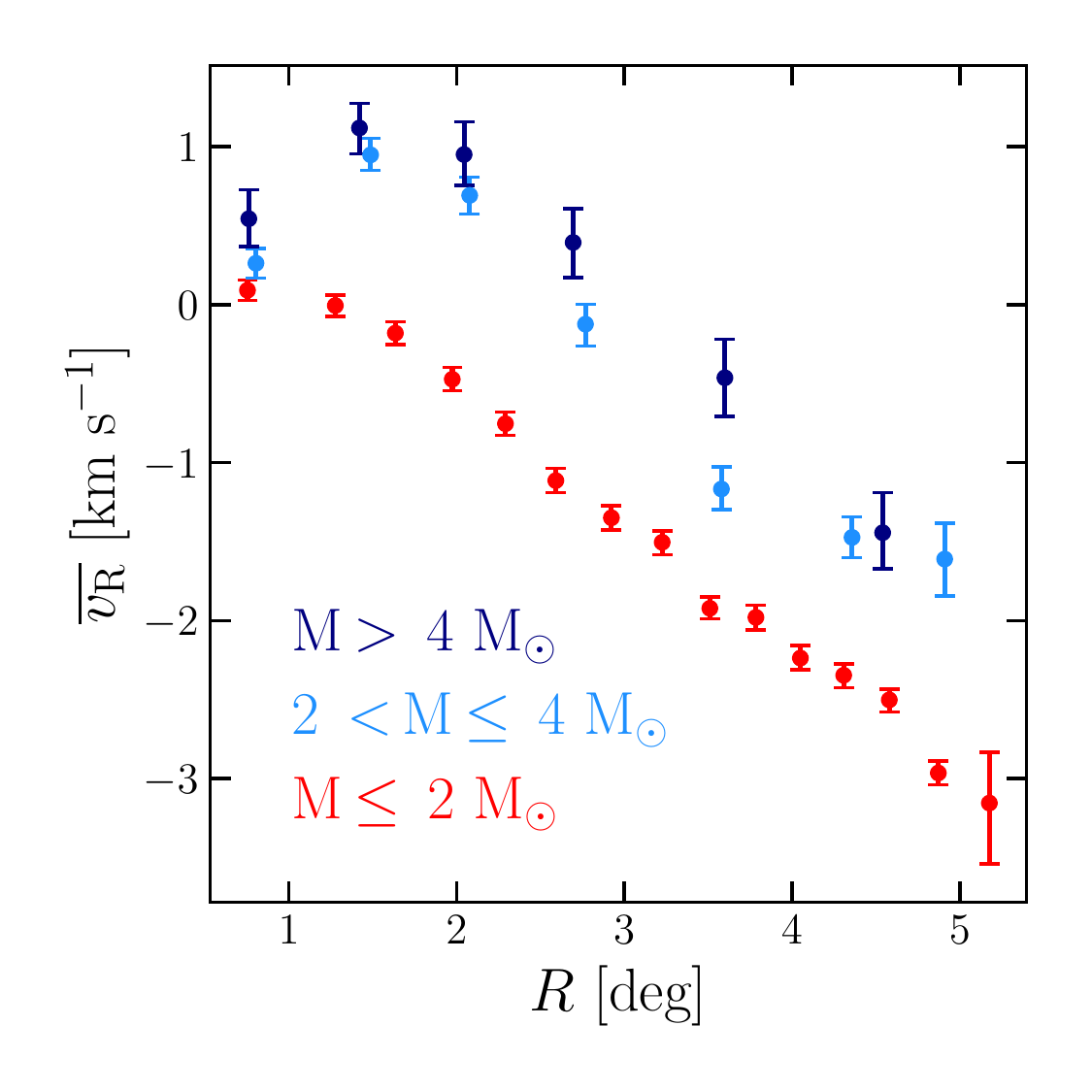}
    \caption{Mean radial velocity profiles for stars in three different mass bins: $M\leq2\,$M$_\odot$ (red markers), $2<M\leq4\,$M$_\odot$ (light-blue markers) and $M>4\,$M$_\odot$ (dark-blue markers). Positive values of $v_{\rm R}$ should be interpreted as expansion and vice versa. \label{fig:expansion_multimass}}
\end{figure}

We also observed a clear dependence of the radial velocity component on the stellar mass.
In particular, stars more massive than about $\gtrsim2\,$M$_\odot$ exhibit a larger positive mean radial velocity (up to about 1 km s$^{-1}$) than lower mass stars within about the innermost 3$^\circ$ (see Figure \ref{fig:expansion_multimass}), while at larger radii massive stars contract toward the center ($\overline{v_{\rm R}} <0$) with a similar slope to lower mass ones albeit with a higher normalization.
\begin{figure*}[h!]
    \centering
    \includegraphics[width=\textwidth]{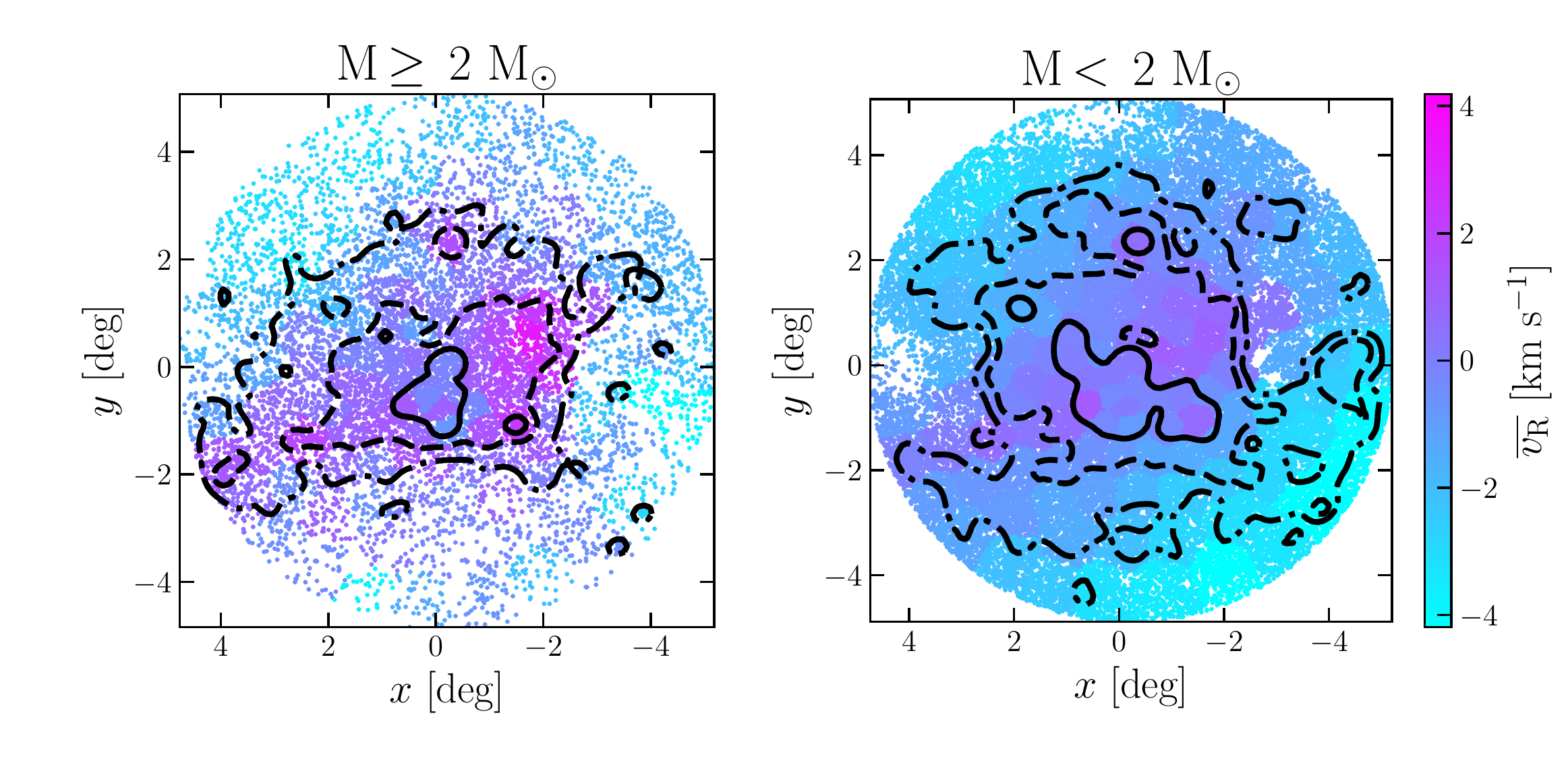}
    \caption{Two-dimensional maps of mean radial velocity for high mass stars ($M\geq2\,$M$_\odot$, left panel) and low mass stars (right panel). Stars belonging to the same Voronoi bin are color-coded according to the mean radial velocity, while black lines are iso-density contours of the respective populations enclosing about $12\%$ (solid line), $39\%$ (dashed line) and $67\%$ (dash-dotted line) of the underlying density distribution. \label{fig:link_density_expansion}}
\end{figure*}

We thus further investigated this feature in the two-dimensional plane using the Voronoi tessellation, giving particular attention to non-spherical-symmetric features and putative links with the spatial distribution of stars. In Figure \ref{fig:link_density_expansion} we show two-dimensional maps of mean radial velocity for stars with mass above (left panel) and below (right panel) $2\,$M$_\odot$. Expansion ($\overline{v_{\rm R}}>0$) is depicted in purple whereas contraction ($\overline{v_{\rm R}}<0$) in blue and black lines highlight iso-probability contours of the density distribution of the respective populations.

Interestingly, a clear connection between expansion and density comes up when looking at massive stars (left panel of Figure \ref{fig:link_density_expansion}), suggesting that higher-density regions expand faster than lower ones, whereas no clear connection is found for the low mass population (right panel).
In addition, we found consistent features to what has been observed in Figure \ref{fig:expansion_multimass}: massive stars expand faster than lower mass ones, which in turn had larger contraction speeds in the outskirts.

The spectro-photometric (i.e. age and metallicity), structural (i.e. density profiles and mass segregation) and kinematical (i.e. contraction and equipartition) evidence collected so far 
suggests that the stars selected within a few degrees from NGC 654 are not just a group of comoving stars but, rather, the nine identified clusters and the extended
low-density halo surrounding them are part of a common, substructured and still assembling massive stellar system.  Following the definition introduced in \citet{dalessandro+2021} we named it LISCA II (where LISCA stands for Lively Infancy of Star Clusters and Associations). 

\section{Structure and kinematic of the embedded star clusters}\label{sec:star_clusters}
In this Section, we present the structural and kinematic properties of the nine star clusters composing LISCA II.

\begin{figure*}[h!]
    \centering
    \includegraphics[width=\textwidth]{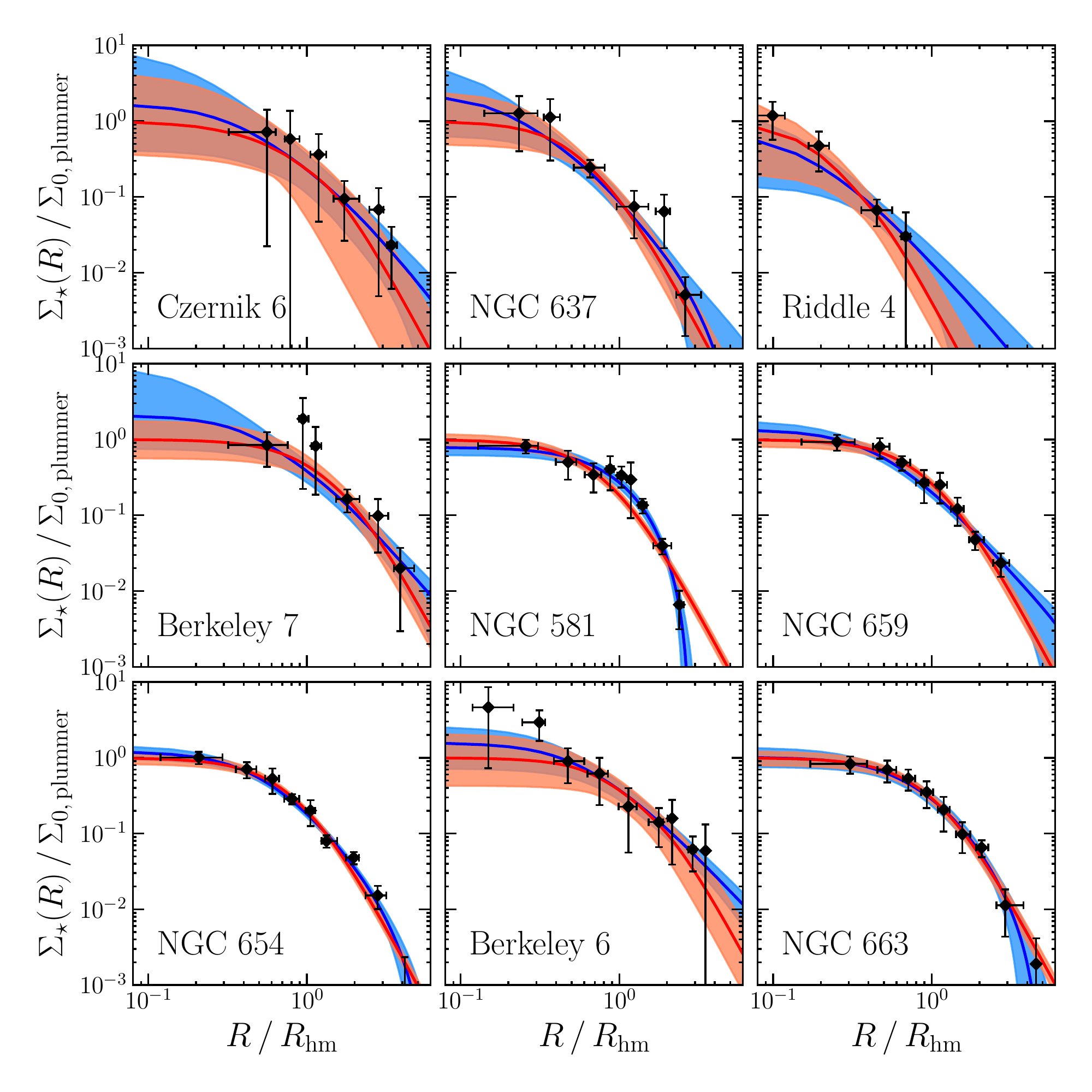}
    \caption{Number density profiles (black markers) for the nine star clusters (names shown in the bottom left corner), along with model predictions for Plummer (red curve) and King (blue curve) models. Shaded areas represent the 68\% credible regions. All the profiles, both observed and models, have been scaled to the central density predicted by a Plummer model $\Sigma_{\rm 0,\,plummer}$ and to the observed projected half-mass radius $R_{\rm hm}$. \label{fig:multipanel_clusterdensity}}
\end{figure*}

First, we constructed the number density radial profiles for all the clusters (following the same approach described in Section \ref{sec:kin_struct}) with respect to their center of mass 
, and we fitted them with Plummer \citep{Plummer1911} and King \citep{king1962} models. 
Figure \ref{fig:multipanel_clusterdensity} shows the observed profiles and the models, normalized to the clusters' central densities and observed projected half-mass radii ($R_{\rm hm}$) enabling a quantitative comparison among different clusters.

The latter quantity is defined as the projected radius which encloses half of the total cluster's mass directly obtained from the radial distribution of member stars.

Every star cluster exhibits a cluster-like profile that, within errors, is equally well modeled by both Plummer and King models, with the only exception of NGC 581 (see Figure \ref{fig:multipanel_clusterdensity}) which exhibits a sharper truncation not captured by a Plummer model.

\begin{figure*}[h!]
    \centering
    \includegraphics[width=\textwidth]{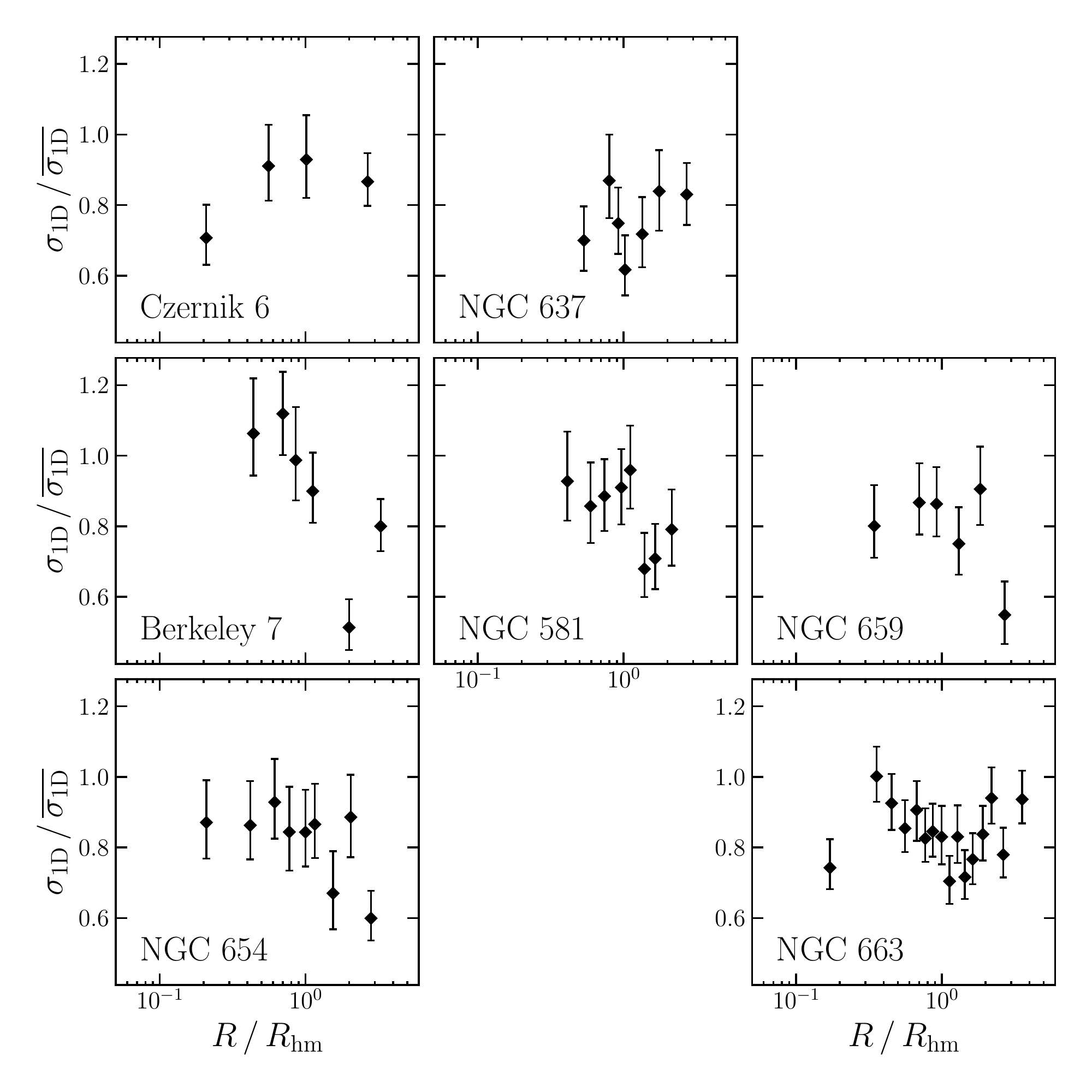}
    \caption{One-dimensional velocity dispersion profiles for seven out of nine star clusters. Velocity dispersion has been normalized to the same quantity computed for all the cluster's members ($\overline{\sigma_{\rm 1D}}$, i.e. without binning the stars) while projected distances from the cluster center of mass have been normalized to the observed half-mass radius. \label{fig:multipanel_clusterdisp}}
\end{figure*}

In addition, we present the kinematic properties of star clusters: Figure \ref{fig:multipanel_clusterdisp} shows the inferred 1D velocity dispersion which has been defined as
\begin{equation}
    \sigma_{\rm 1D} \equiv \sqrt{\frac{\sigma^2_{\rm R} + \sigma^2_{\rm T}}{2}}\,,
    \label{eq:sigma1D}
\end{equation}
with $\sigma_{\rm R}$ and $\sigma_{\rm T}$ being the projected radial and tangential components of the velocity dispersion inferred assuming the likelihood in Equation \ref{eq:likelihood_meandisp} and sampling the parameters' space with an MCMC technique.

For two clusters, namely Riddle 4 and Berkeley 6, we could not compute reliable kinematic profiles since few stars (51 and 84 respectively) fulfilled the astrometric quality selections presented in Section \ref{sec:kin_struct}.

All the other clusters exhibit rather flat dispersion profiles 
that can be hardly explained by equilibrium models, for instance, those adopted to fit the stellar density distributions. Several effects might be at play in producing the observed flat dispersion profiles, such as contaminants from the stellar halo and dynamical heating due to tidal interactions with other clusters and sub-structures possibly taking place during the system's early evolution. We will discuss the latter mechanism in more detail in Section \ref{sec:simulation}.

\begin{table*}
\caption{Star clusters' properties obtained in this study. \label{tab:final_results}}
\centering
\begin{tabular}{lcccccccccc}
\hline\hline
Name & $\alpha_{\rm CM}$ & $\delta_{\rm CM}$ & $\varpi$ & $\mu_{\alpha *}$ & $\mu_{\delta}$ & $a$ & $R_{\rm c}$ & M$_{\rm tot,\,Kroupa}$ & M$_{\rm tot,\,Salpeter}$ & $N_{\rm member}$\\
& [$^\circ$] &[$^\circ$]&[mas]&[mas\,yr$^{-1}$]&[mas\,yr$^{-1}$]&[pc]&[pc]&[10$^2$\,M$_\odot$]&[10$^2$\,M$_\odot$]& \\
(1)&(2)&(3)&(4)&(5)&(6)&(7)&(8)&(9)&(10)&(11)\\
\hline
Czernik 6  & 30.5313 & 62.8293 & 0.335$\pm$0.015 & -1.19$\pm$0.03 & -0.19$\pm$0.06 & $7.4^{+5.3}_{-5.1}$ & $2.8^{+4.0}_{-1.6}$ & 4.9  & 7.7  & 191\\
NGC 637    & 25.8828 & 64.1880 & 0.352$\pm$0.009 & -1.26$\pm$0.04 & -0.03$\pm$0.04 & $5.7^{+2.2}_{-1.9}$ & $1.9^{+2.7}_{-0.9}$ & 10.1 & 15.9 & 291\\
Riddle 4   & 32.0533 & 60.3979 & 0.354$\pm$0.018 & -0.80$\pm$0.04 & -0.50$\pm$0.06 & $2.2^{+1.9}_{-0.7}$ & $1.4^{+2.3}_{-0.5}$ & 4.9  & 7.7  & 67\\
Berkeley 7 & 28.5599 & 62.2259 & 0.340$\pm$0.010 & -0.98$\pm$0.04 & -0.22$\pm$0.04 & $5.8^{+0.6}_{-0.6}$ & $4.0^{+1.5}_{-1.1}$ & 8.9  & 14.1 & 272\\
NGC 581    & 23.3985 & 60.7205 & 0.361$\pm$0.004 & -1.39$\pm$0.04 & -0.60$\pm$0.03 & $3.5^{+0.3}_{-0.3}$ & $4.4^{+1.6}_{-1.1}$ & 14.6 & 23.0 & 319\\
NGC 659    & 26.1205 & 60.6937 & 0.309$\pm$0.008 & -0.83$\pm$0.03 & -0.30$\pm$0.03 & $3.2^{+0.3}_{-0.3}$ & $1.4^{+0.5}_{-0.3}$ & 11.5 & 18.1 & 240\\
NGC 654    & 26.0235 & 61.8833 & 0.325$\pm$0.006 & -1.14$\pm$0.05 & -0.33$\pm$0.04 & $2.3^{+0.2}_{-0.2}$ & $1.2^{+0.2}_{-0.2}$ & 20.8 & 32.8 & 421\\
Berkeley 6 & 27.7991 & 61.0746 & 0.327$\pm$0.005 & -0.90$\pm$0.05 & -0.53$\pm$0.04 & $3.1^{+1.8}_{-1.1}$ & $1.3^{+0.9}_{-0.3}$ & 4.3  & 6.6  & 107\\
NGC 663    & 26.5625 & 61.1930 & 0.341$\pm$0.008 & -1.14$\pm$0.04 & -0.33$\pm$0.04 & $5.8^{+0.6}_{-0.6}$ & $4.0^{+1.5}_{-1.1}$ & 55.8 & 88.0 & 1079\\
\hline
\end{tabular}
\tablefoot{(1) cluster name. (2)$-$(3) center of mass coordinates obtained from stars with mass $M>2\,M_\odot$. (4) mean cluster parallax. (5)$-$(6) mean proper motions. (7) scale length for a Plummer model. (8) core radius of a King model. (9)$-$(10) total system's mass obtained assuming either a Kroupa or a Salpeter IMF. (11) total number of Gaia sources flagged as cluster members.}
\end{table*}

To estimate the total mass of each cluster we followed the same approach described in \citet{dalessandro+2021} and we normalized a \citet[][]{kroupa2001_imf} and a \citet{salpeter1955_imf} initial mass function (IMF) in the range $M>4$ M$_\odot$, such that the number of member stars matched the one predicted by direct integration of the IMF
\begin{equation}
    N_{\star\,{\rm observed}} = \int^{m_{\rm max}}_{m_{\rm min}}\;IMF(m)\,dm\;,
    \label{eq:imf_normalization}
\end{equation}
with $m_{\rm min} = 4$ M$_\odot$ and $m_{\rm max}=11-14$ M$_\odot$ for every cluster but Berkeley 6 whose maximum stellar mass is around 7 M$_\odot$ according to its inferred age.
Once the normalization has been obtained, we computed the total visible mass by integrating the IMF in the range $M_{\rm min}-M_{\rm max} = 0.09-14.73$ M$_\odot$ 
\begin{equation}
     M_{\rm tot} = \int^{M_{\rm max}}_{M_{\rm min}}\;IMF(m)\,m\,\,dm\;,
    \label{eq:imf_totalmass}
\end{equation}
respectively the minimum and maximum stellar mass for a 14 Myr-old simple stellar population with $[{\rm Fe}/{\rm H}] = -0.3$ dex \citep{bressan2012_parsec}.

The derived clusters' masses range between $\simeq 0.5-5.6\times10^3$ M$_\odot$ ($\simeq 0.7-8.8\times10^3$ M$_\odot$) according to a Kroupa (Salpeter) IMF.

Clusters' main kinematic and structural properties are summarized in Table \ref{tab:final_results}.
Specifically, for each cluster, we included: the coordinates of the center of mass (for stars more massive than 2 M$_\odot$), the mean parallax and PM (of the distributions shown in Figure \ref{fig:region_selection}) along with errors, the Plummer scale length and the King core radius (obtained from the models shown in Figure \ref{fig:multipanel_clusterdensity} and converted in parsec using the mean parallax also reported in Table \ref{tab:final_results})
, the total masses inferred by assuming either a Kroupa or a Salpeter IMF, and finally the total number of member stars for each cluster.

\section{Total system's mass}\label{sec:total_system_mass}
The total mass of LISCA~II has been estimated by using equation \ref{eq:imf_totalmass} and roughly the same approach as for the single clusters. However, in this case, we had to
assume a radial extension of the system within which to integrate the stellar masses. To this aim we
used the Jacobi radius \citep[$R_{\rm J}$ - see e.g.][]{binneyTremaine2008} as a first-order
physically plausible radial extension of the system. $R_{\rm J}$ is simply defined as follows
\begin{equation}
    R_{\rm J} = R_0\,\left( \frac{M_{\rm LISCAII}}{3M_{\rm MW}(<R_0)}\right)^{1/3}\,,
    \label{eq:jacobi_radius}
\end{equation}
with $R_0\simeq10.2$ kpc 
and $M_{\rm MW}(<R_0)= 10.98^{+0.12}_{-0.10}\times10^{10}$ M$_\odot$ being the Galactocentric distance of the system and the MW mass enclosed within that radius, respectively \citep{Cautun+2020}, and $M_{\rm LISCAII}$ is the total system's mass.
Starting from Equation \ref{eq:jacobi_radius}, we estimated both $R_{\rm J}$ and the system's mass by using an iterative procedure until final convergence was reached.
Depending on the assumed IMF we obtain:
\begin{eqnarray}
    R_{\rm J,\,Kroupa} = 1.07\,^\circ \simeq 55\,{\rm pc} \,,\nonumber\\
    R_{\rm J,\,Salpeter} = 1.35\,^\circ \simeq 70\,{\rm pc}\,,
    \label{eq:jacobi_radius_LISCA2}
\end{eqnarray}
with the corresponding enclosed ($<2\,R_{\rm J}$) masses being
\begin{eqnarray}
    M_{\rm LISCAII,\,Kroupa} = 6.4\times10^4 M_\odot \,,\nonumber\\
    M_{\rm LISCAII,\,Salpeter} = 1.2\times10^5 M_\odot\,.
    \label{eq:totalmass_halo_LISCA2}
\end{eqnarray}
It is important to emphasize here that in this case $R_{\rm J}$ is only meant to provide a general indication of the spatial scale related to the strength of the tidal field at the location of LISCA~II. Indeed, in the complex case of a clumpy system far from a spherical configuration in dynamical equilibrium like that of LISCA~II, the detailed implications of the effects of the tidal field truncation for this system would require a tailored set of simulations. We also point out that the kinematical properties revealed by our analysis (Section \ref{sec:kin_struct}) indicate that in this system all the stars, including those beyond the present estimate of $R_{\rm J}$ are strongly contracting towards the center of the system, which is just the opposite of what is expected for stars escaping the system.

\section{Comparison with $N$-body simulations of early cluster evolution}\label{sec:simulation}

\begin{figure}[h!]
    \centering
    \includegraphics[width=0.5\textwidth]{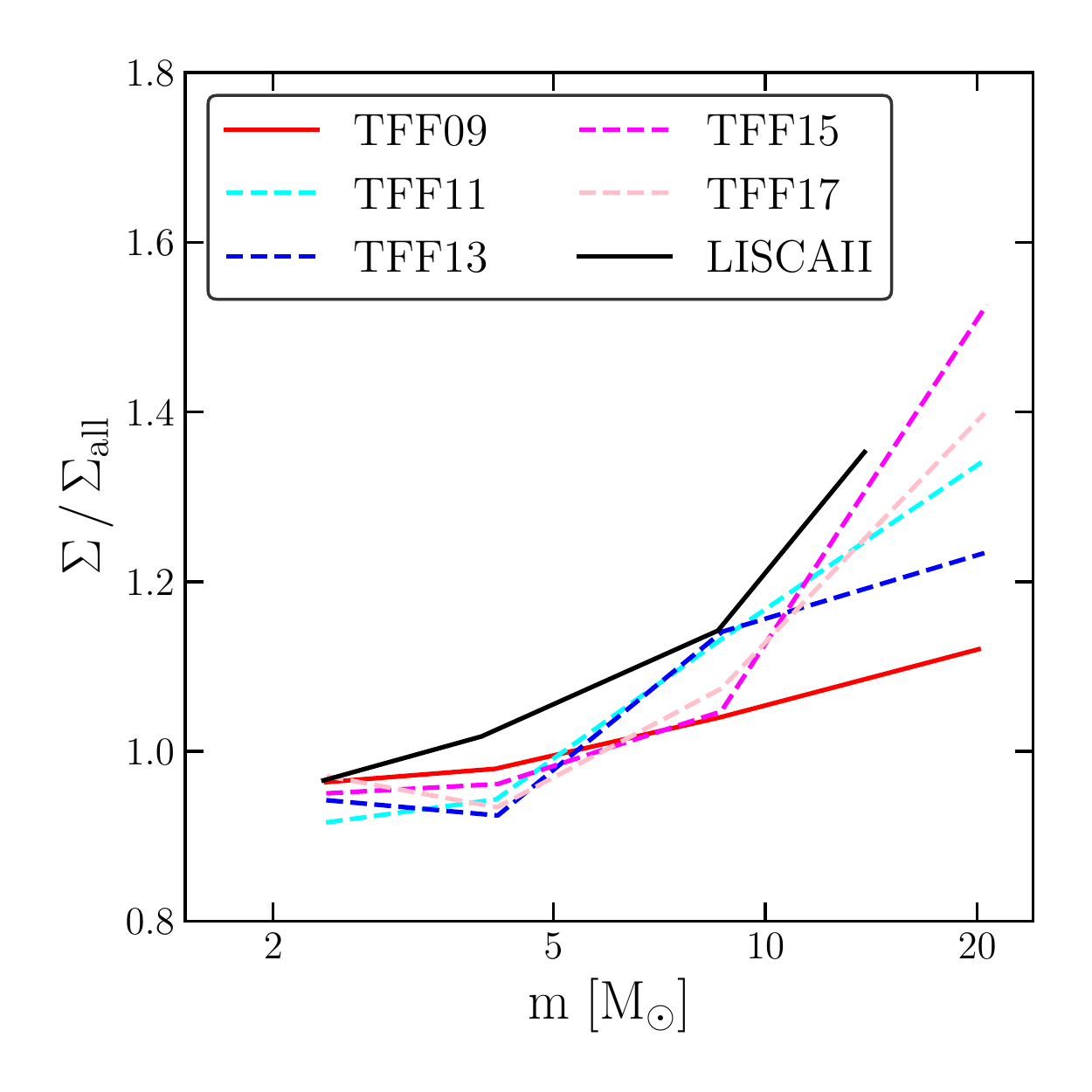}
    \caption{Local number density of stars as a function of stellar mass (see Section \ref{sec:sim_massseg_intmotion}). The local density is normalized by the median local surface density of the entire cluster. The increase of local density with the stellar mass indicates  mass segregation on local scales across all clusters shown. \label{fig:ldmseg}}
\end{figure}
\begin{figure}[h!]
    \centering
    \includegraphics[width=0.5\textwidth]{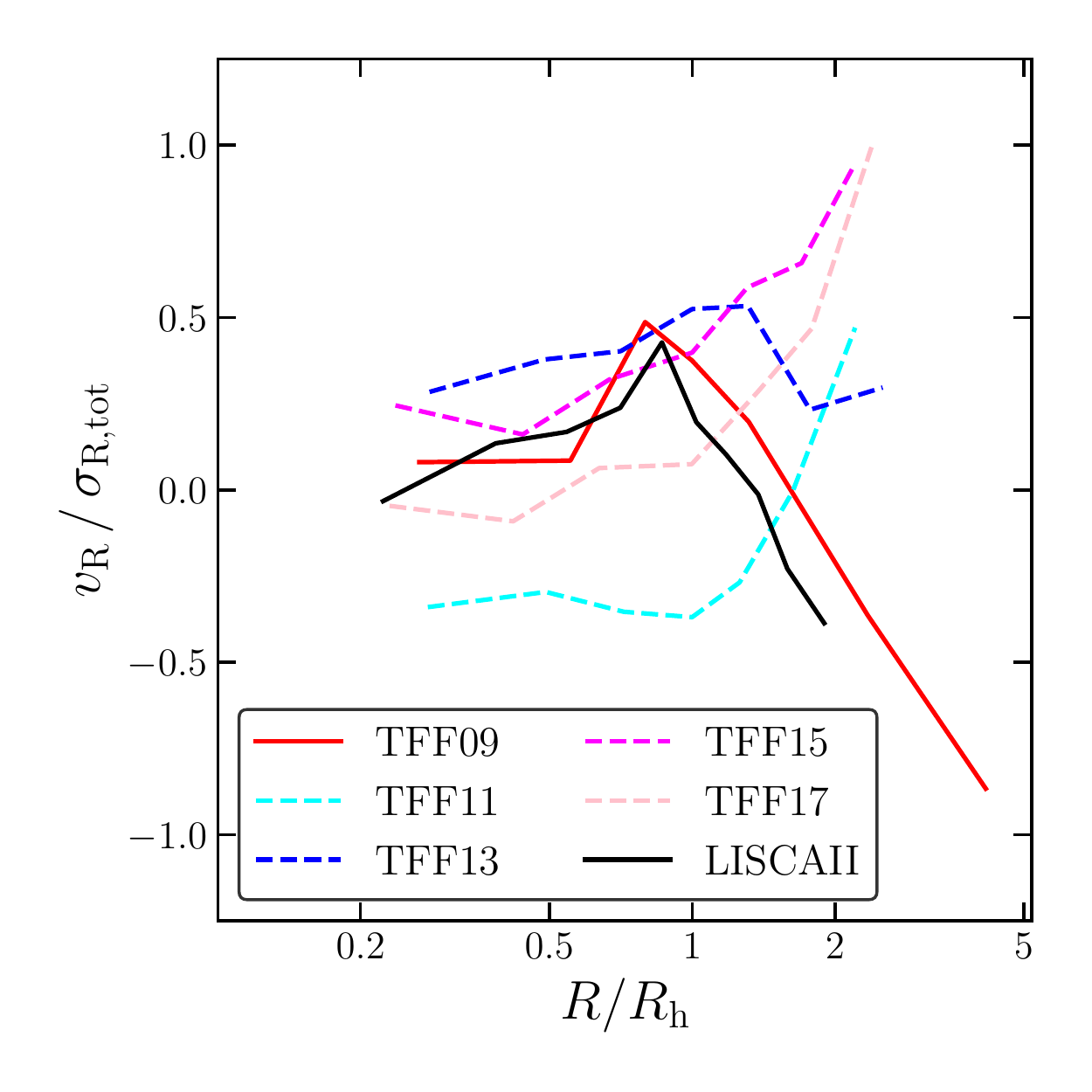}
    \caption{Projected radial velocity profiles for each snapshot of the $N$-body simulation and the LISCA II system. The projected radius has been normalized to the projected half-mass radius of the system in each snapshot, and the radial velocity is normalized to the radial velocity dispersion of all stars in the system. \label{fig:gvp}}
\end{figure}

\begin{figure}[h!]
    \centering
    \includegraphics[width=0.5\textwidth]{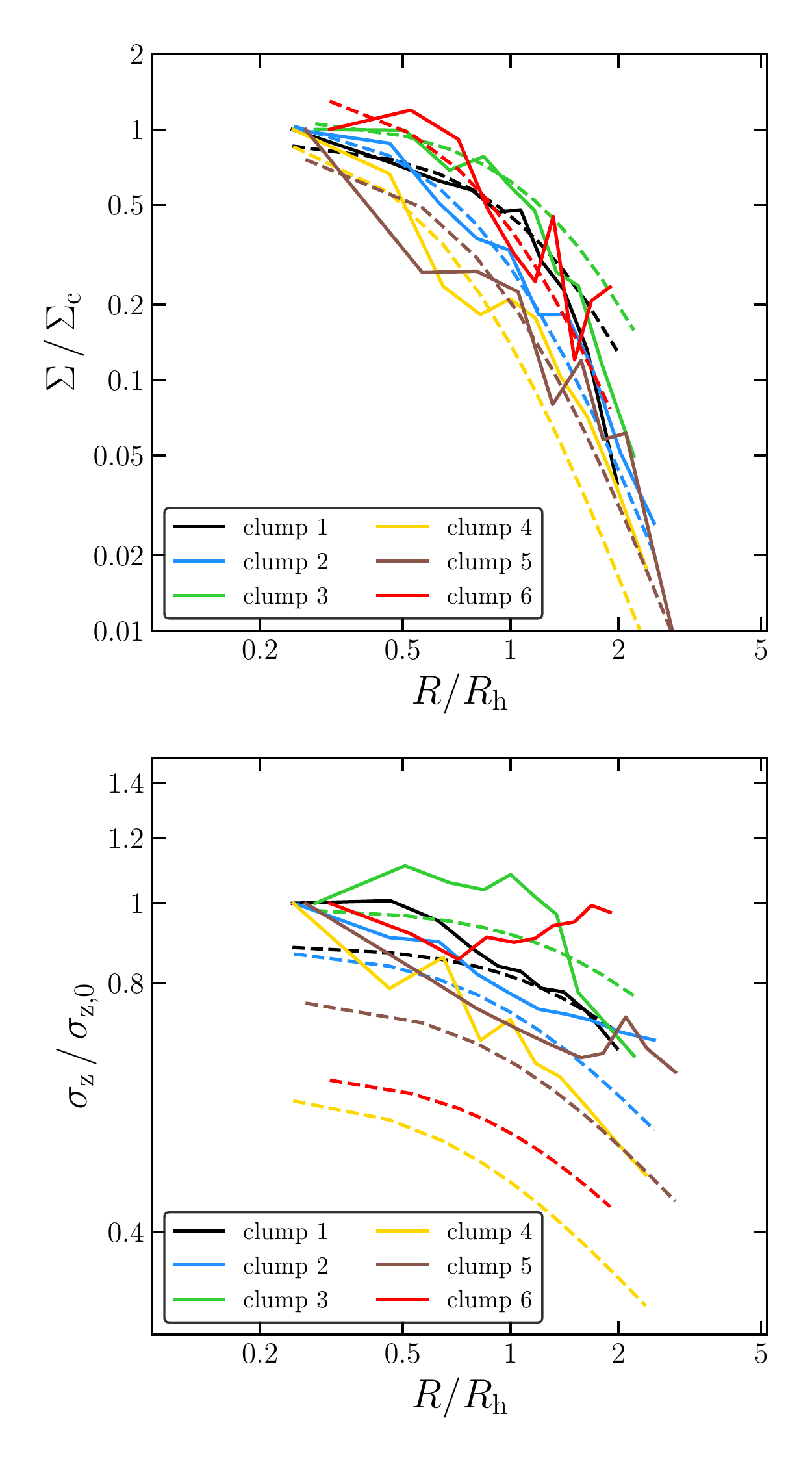}
    \caption{Surface density profiles (top) and one-dimensional velocity dispersion profiles (bottom) for the TFF09 snapshot. Best-fit Plummer models are also shown, indicated by dashed lines for each clump. The best-fit Plummer model is determined by fitting  the surface density profile within one projected half-mass radius. The surface density and one-dimensional velocity dispersion are both normalized to the corresponding values in innermost radial bin of each clump. \label{fig:densdispfits}}
\end{figure}
\begin{figure}[h!]
    \centering
    \includegraphics[width=0.5\textwidth]{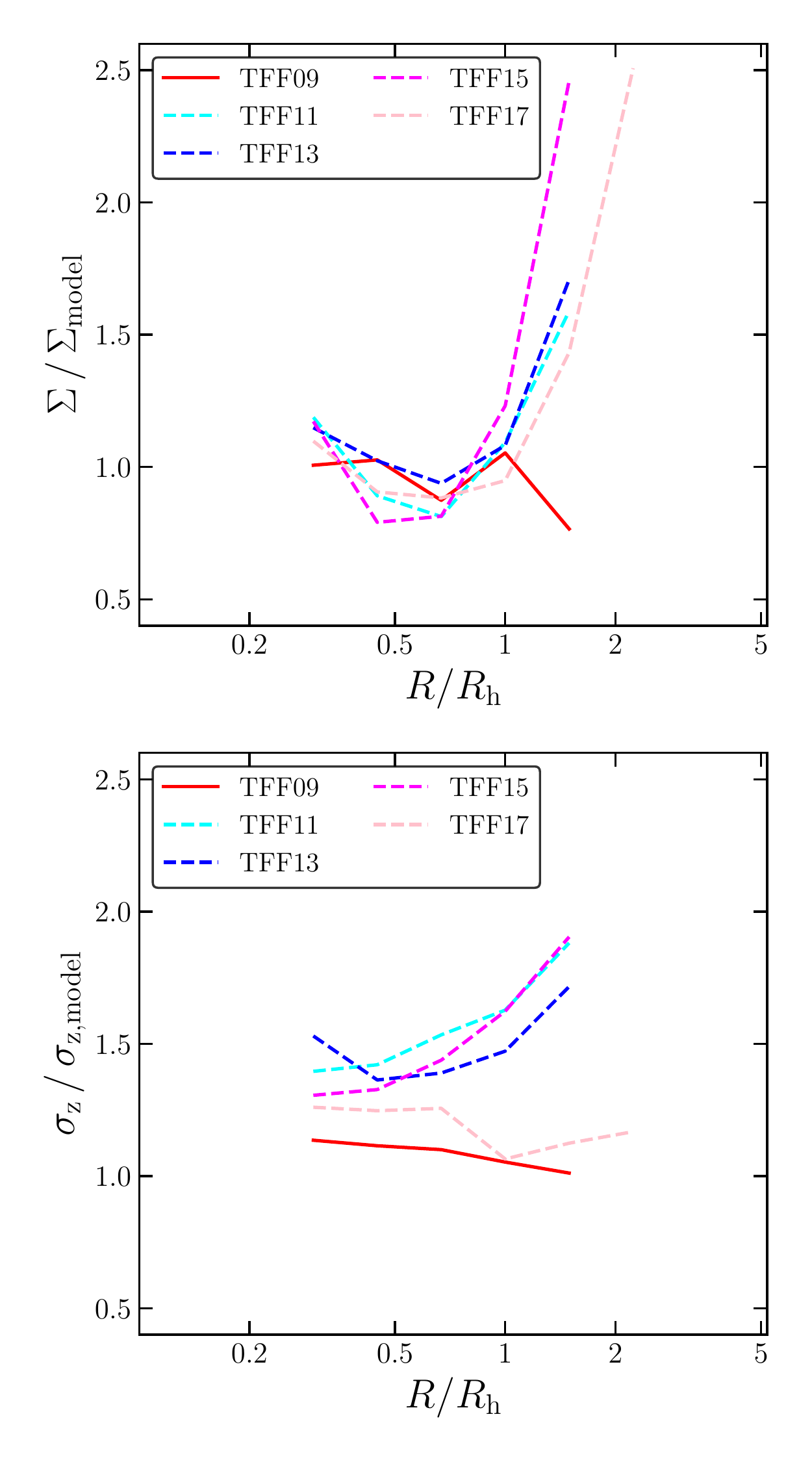}
    \caption{Radial profiles of the ratio between the clump radial profiles and the best-fit Plummer models for the surface density (top) and one-dimensional velocity dispersion (bottom). This ratio is averaged over the three largest clumps within each snapshot. \label{fig:densdispmodel}}
\end{figure}

\begin{figure}[h!]
    \centering
    \includegraphics[width=0.5\textwidth]{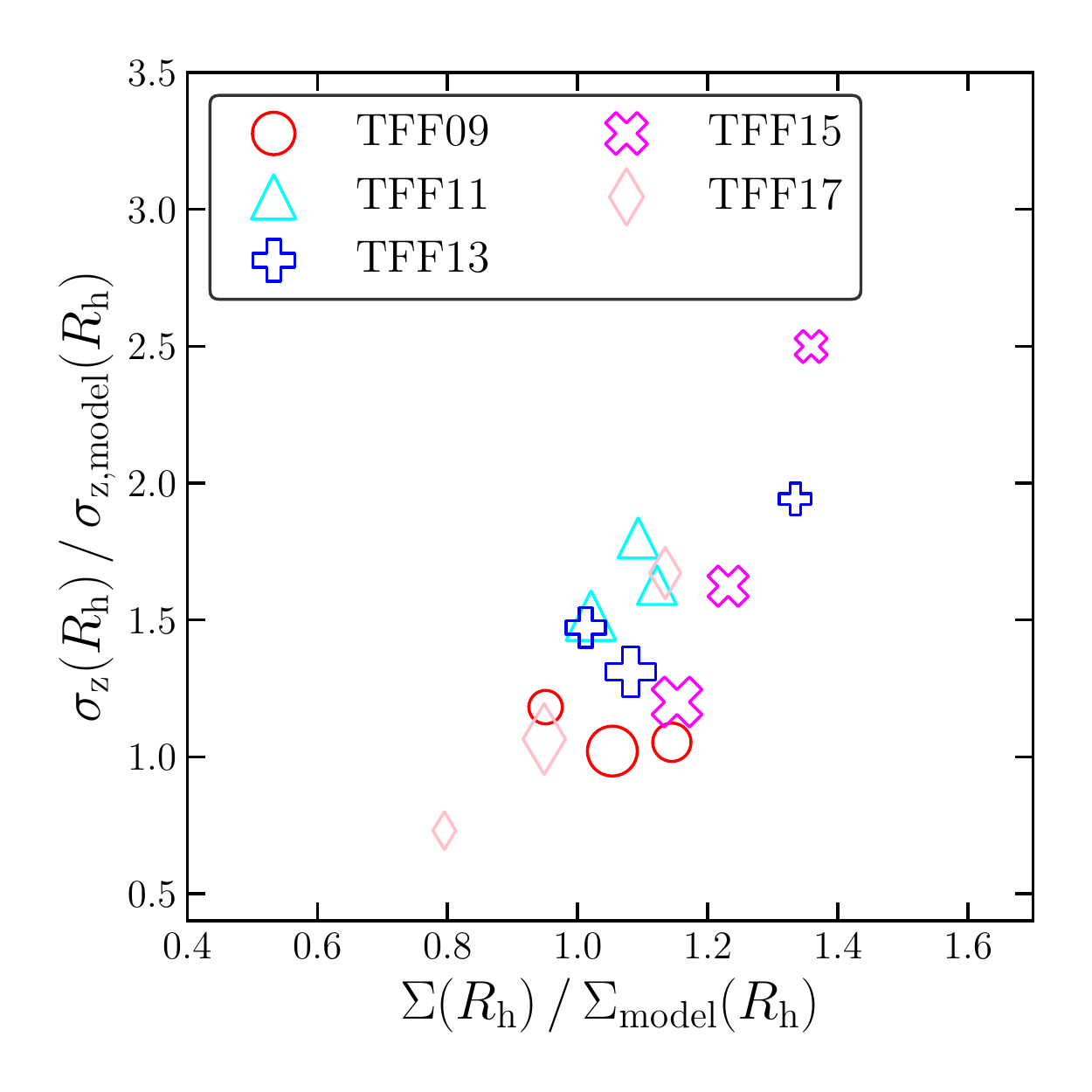}
    \caption{One-dimensional velocity dispersion versus the projected surface density for the three largest clumps in each snapshot shown in the legend, each calculated at one projected half-mass radius. Both the one-dimensional velocity dispersion and the projected surface density are normalized by the value expected from the Plummer model best fitting the surface density profile. Point sizes are proportional to the total mass of each clump.\label{fig:vzratdens}}  
\end{figure}
In this Section, we present an analysis of the dynamical properties of one of the $N$-body models studied in \citet[][]{livernois+2021}. The simulations of \citet[][]{livernois+2021} explored the early evolution and violent relaxation phases of  young rotating star clusters and followed their evolution from a hierarchical structure to a final monolithic equilibrium configuration.

We emphasize that our goal here is not to build a detailed model of the LISCA II system but rather to gather further insight into the interpretation of our observational analysis by providing a theoretical example of the general dynamical properties expected in a hierarchical stellar cluster undergoing its early evolutionary phases.

The model we analyze for this paper is the F025 model of \citet[][]{livernois+2021}. The model is fully described in \citet[][]{livernois+2021}, but we summarize its main features for the purposes of this study. The model starts with $10^5$ stars with mass range $0.08-100$ M$_\odot$, initially following a fractal distribution with a fractal dimension equal to 2.6; the system is initially dynamically cold and undergoes the collapse and subsequent structural oscillations typical of the violent relaxation phase.
As a reference timescale for the presentation of our results, we adopt the system free-fall timescale, $t_{\rm ff}$, which corresponds approximately to the timescale needed for the system to reach its maximum contraction during its initial collapse. 
Assuming an initial mass of $5\times10^4 - 10^5$ M$_{\odot}$ and an initial radius of $\sim50 - 60$ pc, the $t_{\rm ff}$ would grossly correspond to $20-35$ Myr.
To capture the model at multiple evolutionary stages, we focus our attention on the snapshots at the following values of $t/t_{\rm ff}$: 0.9 (denoted TFF09), 1.1 (TFF11), 1.3 (TFF13), 1.5 (TFF15), and 1.7 (TFF17).

\subsection{Mass segregation and bulk internal motion}
\label{sec:sim_massseg_intmotion}
We start our analysis with the study of mass segregation in the $N$-body model and the LISCA II system. Here, we focus our attention on an analysis specifically aimed at detecting mass segregation on a local scale which might provide further insight into the dynamics of systems characterized by the presence of clumps and substructures like those studied here. 
To quantify the level of mass segregation, we first calculate the local surface number density for each star using the distance of the sixth nearest star of any mass \citep{casertano&hut_1985}; the median local surface density of different mass bins, normalized by the median local surface density of the entire cluster, is plotted against the median stellar mass of each bin in Figure \ref{fig:ldmseg}.
All snapshots show clear evidence of a local surface density increasing with the stellar mass, implying the presence of local-scale mass segregation, where massive stars are migrating toward the centers of the sub-clusters they are members of. A similar trend is present also in LISCA II which shows significant local-scale mass segregation, complementing the global-scale mass segregation found in Figure \ref{fig:cumulative_profiles}. This trend appears to evolve with time in the model snapshots, with the two latest snapshots having the strongest trend of local-scale mass segregation.

Focusing on the bulk internal motion of the system, we analyze the radial velocity profile of the simulation data as a comparison to Figure \ref{fig:radial_contraction_profiles}. 
In Figure \ref{fig:gvp}, we plot the radial velocity profile normalized by the velocity dispersion of the cluster, including only stars with $m>2M_{\odot}$. 
As the cluster evolves, we see the different regions of the cluster transition between expansion and contraction. The TFF09 snapshot is characterized by a trend similar to that found in LISCA II: an expansion of the inner regions and a strong contraction in the outer regions.

\subsection{Dynamics of sub-clusters}
For insight into the possible dynamical evolution of the sub-clusters within LISCA II, we analyzed the dynamical properties of selected clumps from our $N$-body models.
For each snapshot in the $N$-body data, we have selected a few clumps in spherical 3-D regions and determined their centers as the location of the maximum local density. These clumps are in dynamically active environments, and no clustering metrics were found to be appropriate across all snapshots for clump identification.

We start by showing in Figure \ref{fig:densdispfits} the surface density profile (top panel) and line-of-sight velocity dispersion profile (bottom panel) for the clumps selected in the TFF09 snapshot.
We fit a Plummer model based on the surface density profile and enclosed mass within one projected half-mass radius, and over plot the best-fit model lines. 
All clumps show radial variation of the surface density profiles following the general shape of the Plummer model; on the other hand, the velocity dispersion profiles are  flatter and more elevated than is expected from the best-fit Plummer models. This is the manifestation of the tidal heating in the cluster environment and is similar to what is seen in Figure \ref{fig:multipanel_clusterdisp}.

We have repeated the above analysis for all of our snapshots, and have plotted in Figure \ref{fig:densdispmodel} the average ratio of the surface density (top panel) and line-of-sight velocity dispersion (bottom panel) to the best-fit Plummer models across the 3 biggest clumps in each snapshot. 
The surface density fits well out to around 1 projected half mass radius, outside of which the clumps generally have a higher density than the best-fit Plummer model; these deviations from the best-fit Plummer model can be attributed to the high density of the surrounding environment and the perturbations due to interactions in the cluster environment. 
The effects of these perturbations are clearly visible in the line-of-sight velocity dispersion profiles of all the snapshots analyzed: all the velocity dispersion profiles deviate from the profiles expected from the best-fit Plummer models across all radii with a dependence on the projected radius that varies in different snapshots.

Finally, we summarize the relation between structural and kinematic perturbations by plotting, in Figure \ref{fig:vzratdens}, the ratio of density and line-of-sight velocity dispersion to the corresponding values of these quantities from the best-fit Plummer models at $R_{\rm h}$ for the three biggest clumps in each snapshot. 
This plot clearly shows that the fingerprints of the highly active environment where each clump is undergoing rapid interactions, mergers, and fragmentations, are more evident in the kinematic properties, as illustrated by the larger deviations of the velocity dispersion from the expected equilibrium values.

\section{Discussion and Conclusions} \label{sec:conclusion}
The unprecedented quality of Gaia DR3 data \citep{gaiaDR3_2022}, supplemented by high-resolution spectra \citep{fanelli_etal2022} obtained as part of the SPA-TNG large program, allowed us to identify the LISCA~II system in the Perseus complex.

The spectro-photometric, structural, and kinematical properties of this system are in generally good agreement with those theoretically expected from the early dynamical evolution of a massive molecular cloud that experienced violent relaxation and is now in the process of hierarchically assembling its stellar constituents and evolving toward a monolithic structure. 
In particular, the observed evidence of mass segregation on a local and global scale, the mass-dependent kinematic properties, and out-of-equilibrium internal kinematics of individual sub-clusters as well as the observation of a dominant contraction pattern toward the system center mainly driven by the external regions of LISCA II and of a milder central expansion, are compatible with what expected in the early evolutionary phases of stellar systems assembling as a coherent massive structure by $N$-body models. The properties of these hierarchical stellar systems are shaped by a combination of large-scale variations of the system's potential and smaller scale interactions of individual sub-clusters and clumps. 

Although more detailed models and additional data would be necessary to further explore the possible fate of this system, the evidence collected in this paper suggests that LISCA II is a good candidate to evolve into a young, massive ($10^4-10^5$ M$_\odot$) star cluster in a timescale of $\sim100$ Myr, corresponding to a few free-fall times.
These results make LISCA~II the second structure, after LISCA~I \citep{dalessandro+2021}, ever found in the MW in the process of hierarchically assembling in a massive stellar cluster. LISCA~II is located at only $\sim 6$ deg from LISCA~I, with which it shares similar chemical composition ([Fe/H]$=-0.30$ dex), age (t$\sim20$Myr) and overall mass.

In conclusion, the present analysis has provided a comprehensive characterization of the process of cluster assembly with a level of detail that cannot be achieved in external galaxies or at high redshift (where the progenitors of the oldest clusters formed), thus showing that, probing cluster formation in local environments 
can help shed light on the physical processes involved in massive cluster formation and their role in determining the cluster's dynamical properties.
Moreover, we further showed that hierarchical cluster assembly is a viable process also in low-density environments, such as the MW \citep[former observational evidence was mainly in high-density environments, e.g.][]{bastian+11,chandar+11} and a statistical assessment of its effectiveness on Galactic scales is the subject of an undergoing study.
It is interesting to note in this respect that the possible observed internal age spreads of the stellar populations belonging to LISCA~I and II ($\sim10$ Myr) nicely fit the observed trend \citep{parmentier+14} between the cluster formation environment stellar density 
and final cluster age internal variations, which possibly results from the different duration of the star formation processes and the link between their efficiency and the systems' free-fall time. This further strengthens the idea that clusters with different present-day properties likely underwent similar formation processes.

\begin{acknowledgements}
The authors thank the anonymous referee for the careful reading of the paper and the useful comments and suggestions.
A.D.C., E.D., and L.O. acknowledge financial support from the project Light-on-Dark granted by MIUR through PRIN2017- 2017K7REXT contract.
E.D. acknowledges support from the Indiana University Institute for Advanced Study through the Visiting Fellowship program.	
This work uses data from the European Space Agency (ESA) space mission Gaia. Gaia data are being processed by the Gaia Data Processing and Analysis Consortium (DPAC). Funding for the DPAC is provided by national institutions, in particular, the institutions participating in the Gaia Multi-Lateral Agreement (MLA). 
\end{acknowledgements}

\appendix
\section{Completeness distributions in different mass bins \label{appendix:completeness}}
\begin{figure}[h!]
    \centering
    \includegraphics[width=0.5\textwidth]{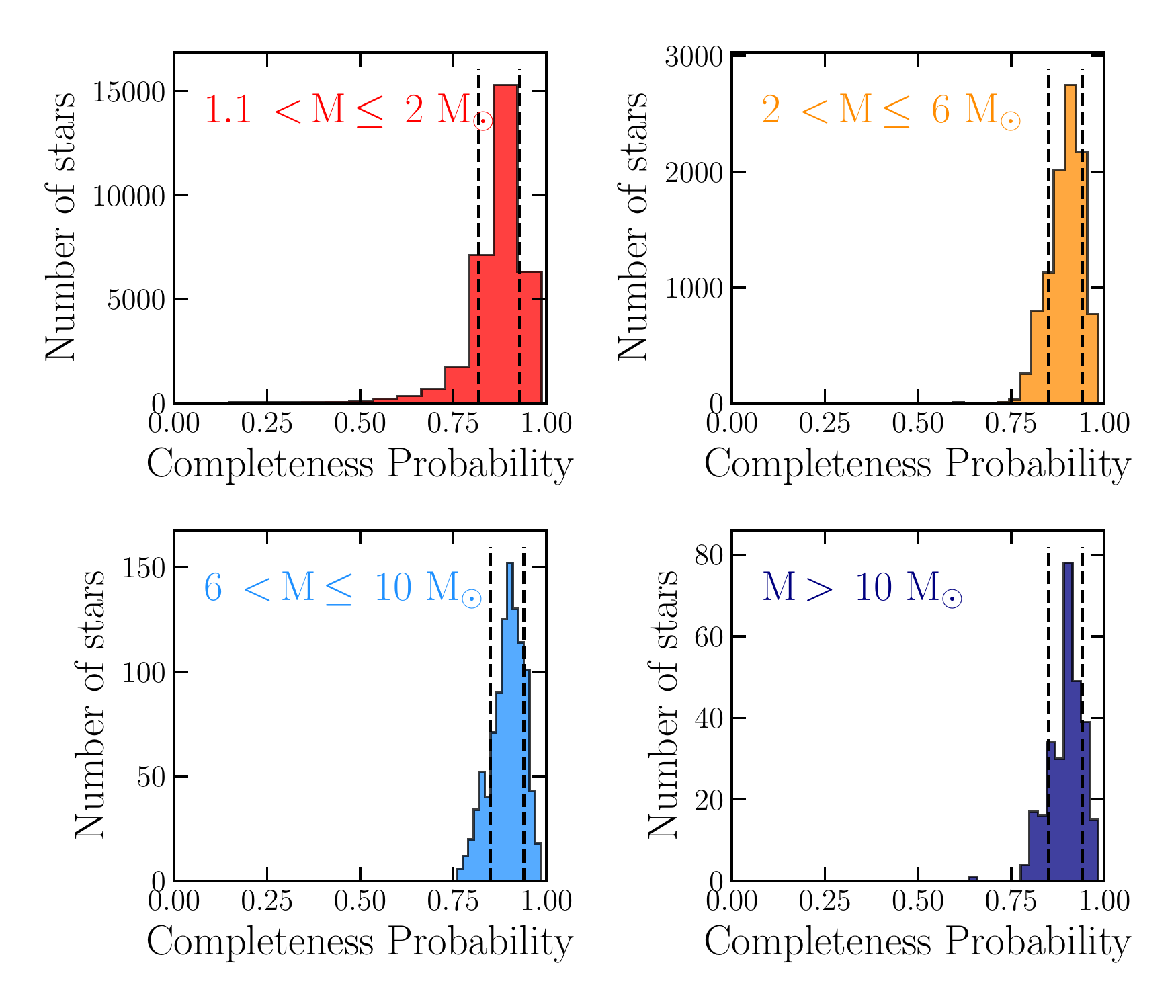}
    \caption{Completeness distribution in four mass bins. Colors are consistent with Figure \ref{fig:cumulative_profiles} and the mass range is shown in the upper left corner.}
\end{figure}
In this Appendix, we show the completeness distributions for the 5 parameter solutions in the four different mass bins used to investigate the presence of mass segregation (see Figure \ref{fig:cumulative_profiles}).
As could be seen, the higher the mass the higher the completeness. The distribution width is due to both a finite mass range in the bin and differential reddening which link the estimated mass to the observed $G$ magnitude.

\bibliographystyle{aa}
\bibliography{bibliography_paper}

\end{document}